\documentclass[journal]{IEEEtran}
\ifCLASSINFOpdf
\else
\fi

\usepackage{graphicx}
\usepackage{amssymb,amsmath,epsfig}
\usepackage{latexsym}
\usepackage{graphicx,multirow,threeparttable,float,balance,array,bm,subcaption,caption}
\usepackage{cite}
\usepackage{hyperref}
\usepackage[english]{babel}
\usepackage{diagbox}
\usepackage{color}

\makeatletter
\begin{document}
\title{Target Speaker Verification with Selective Auditory Attention for Single and Multi-talker Speech}
\author{Chenglin~Xu,~\IEEEmembership{Member,~IEEE,}
Wei~Rao,~\IEEEmembership{Member,~IEEE,}
Jibin~Wu,~\IEEEmembership{Member,~IEEE,}
and~Haizhou~Li,~\IEEEmembership{Fellow,~IEEE}% <-this % stops a space
\thanks{
Submitted to IEEE/ACM transaction on Audio, Speech and Language on 10 Jan. 2021.
This research/project is supported by the National Research Foundation, Singapore under its AI Singapore Programme (AISG Award No: AISG-100E-2018-006); Human-Robot Interaction Phase 1 (Grant No. 192 25 00054), National Research Foundation (NRF) Singapore under the National Robotics Programme (\textit{Corresponding author: Wei Rao}).\\
Chenglin Xu, Jibin Wu and Haizhou Li are with the Department of Electrical and Computer Engineering, National University of Singapore, Singapore (e-mail:\{elexucl, elejbw, haizhou.li\}@nus.edu.sg). Wei Rao is with Tencent Ethereal Audio Lab, Shenzhen, China (e-mail:ellenwrao@tencent.com).
}}

%\thanks{Manuscript received April 1, 2015.}}

%\markboth{Journal of IEEE Audio, Speech and Language Processing Magazine, April~2015}%
%{Shell \MakeLowercase{\textit{et al.}}: Bare Demo of IEEEtran.cls for Journals}



\maketitle

\begin{abstract}

Speaker verification has been studied mostly under the single-talker condition. It is adversely affected in the presence of interference speakers. Inspired by the study on target speaker extraction, e.g., SpEx, we propose a unified speaker verification framework for both single- and multi-talker speech, that is able to pay selective auditory attention to the target speaker. This target speaker verification (tSV) framework jointly optimizes a speaker attention module and a speaker representation module via multi-task learning. We study four different target speaker embedding schemes under the tSV framework. The experimental results show that all four target speaker embedding schemes significantly outperform other competitive solutions for multi-talker speech. Notably, the best tSV speaker embedding scheme achieves 76.0\% and 55.3\% relative improvements over the baseline system on the WSJ0-2mix-extr and Libri2Mix corpora in terms of equal-error-rate for 2-talker speech, while the performance of tSV for single-talker speech is on par with that of traditional speaker verification system, that is trained and evaluated under the same single-talker condition.

\end{abstract}


% \fi
\IEEEpeerreviewmaketitle

\section{Introduction}
\label{sec:introduction}

\IEEEPARstart{T}{}raditional speaker verification (SV) methods, such as i-vector \cite{Dehak&Kenny2011, Kenny10, Garcia11} with probabilistic linear discriminant analysis (PLDA) \cite{prince2007probabilistic}, x-vector PLDA \cite{snyder2016deep,snyder2018x,snyder2019speaker}, assume that input speech is uttered by a single speaker. These methods, however, degrade significantly in the presence of interference speakers. %, even though the multi-talker speech is included during training. 
Speaker diarization technique seeks to inform `who spoke when?' It segments the multi-talker speech temporally into speaker turns, and identifies speaker-overlapping segments %Speaker diarization is typically used to solve the multi-talker speaker verification problem when the multi-talker speech is non-overlapped
\cite{anguera2012speaker,kudashev2016speaker,liu2016sitw,novotny2016analysis,ghaemmaghami2016speakers,Snyder2019multi-speaker}. By doing so, one is able  to exclude speaker-overlapping segments from speaker verification~\cite{charlet2013impact,yella2014overlapping}. Along the same line of thought, the recent studies on target-speaker  voice activity detection (VAD) show that we are able to obtain the target speaker's boundary in a multi-talker speech, e.g. personal VAD \cite{ding2019personal}, target VAD \cite{medennikov2020target}.  In general, the speaker diarization technique is helpful only if the speakers overlap sporadically, while it fails when the speakers are heavily overlapped in time. 

%However, the personal or target VAD only provides the time boundary of the target speaker. If the target speaker's voice is overlapped with an interference speaker, the interference speaker's voice is not removed and this overlapped segment is still included into the target speaker's time boundary. Therefore, the persoanal VAD or target VAD doesn't significantly improve the performance of the SV systems in heavy overlapped multi-talker environment.

From the time-frequency analysis point of view,  multi-talker speech can be considered as multiple single-talker speech samples overlapping in both temporal and spectral dimensions. To recover a single-talker speech sample, the monaural speech separation techniques could come in handy. Successful implementations include deep clustering \cite{hershey2016deep}, deep attractor network \cite{chen2017deep}, permutation invariant training \cite{kolbaek2017multitalker, xu2018single, xu2018shifted}, Conv-TasNet \cite{luo2019conv}, DPRNN \cite{luo2019dual}.  However, speech separation technique seeks to recover the single-talker speech for each individual, that is not only an overkill for speaker verification, but also difficult particularly when we don't know the number of speakers in the multi-talker speech. 
%While speech separation techniques have advanced by leaps and bounds, the number of speakers has to be known in prior for the speech separation techniques to work well. However, the number of speakers in the test speech is always unknown that limits many real-world applications. Besides this, the complexity of the SV system is linearly increased along with the increase of separated sources, because each separated source needs to be confirmed whether this separated source is the enrolled speaker.  In the SV task, we are only interested in the target speaker's voice.

For speaker verification, we are only interested in the presence or absence of the target speaker. Speaker extraction technique, a variant of speech separation, that aims to extract one target speaker at a time, is clearly more relevant to the speaker verification task. The question is how to optimize the speaker extraction algorithm so as to better serve the purpose of speaker verification. The speaker extraction techniques typically rely on a reference speech to direct the selective auditory attention to the target speaker in the observed speech~\cite{xu2020spex}. For brevity, the selective auditory attention mechanism is referred to as speaker attention hereafter. Many successful speaker extraction techniques are proposed recently, instances include SpeakerBeam \cite{delcroix2018single}, SBF-MTSAL \cite{xu2019optimization}, SBF-MTSAL-Concat \cite{xu2019optimization}, Voicefilter \cite{wang2019voicefilter}, DENet \cite{wang2018deep}, SpEx \cite{xu2020spex}, and SpEx+~\cite{ge2020spex+}. In speaker verification, a reference speech of the target speaker is always available, that is required by the enrollment process, thus also called enrollment utterance. Such reference speech can be readily used to direct speaker attention. 

%we always compare the test speech sample with a reference speech of the target speakerAs the SV task always provides a reference speech of the target speaker, known as the enrollment utterance, recent speaker extraction techniques could be a suitable approach to perfectly solve the overlapped multi-talker SV problem. Speaker extraction takes the reference of the target speaker to form an attention and only extracts the speech of this target speaker from the overlapped multi-talker mixture, such as, SpeakerBeam \cite{delcroix2018single}, SBF-MTSAL \cite{xu2019optimization}, SBF-MTSAL-Concat \cite{xu2019optimization}, Voicefilter \cite{wang2019voicefilter}, DENet \cite{wang2018deep}, SpEx \cite{xu2020spex}, SpEx+ \cite{ge2020spex+}, and so on.

The idea of speaker extraction (SE) followed by speaker verification, i.e., SE-SV \cite{rao2019target_is} pipeline, was previously studied to address speaker verification for multi-talker speech. The SE-SV system extracts the speech of the target speaker in the first stage, and subsequently processes the extracted speech with a standard speaker verification module, such as i-vector PLDA~\cite{Dehak&Kenny2011, Kenny10, Garcia11}. %The frequency-domain SBF-MTSAL and SBF-MTSAL-Concat methods are studied as the target speaker extraction modules, respectively. 
Unfortunately, within the SE-SV framework, the speaker extraction front-end and the speaker verification back-end are optimized separately, leading to a potential mismatch between the two modules. Furthermore, the SE-SV system is designed for overlapped multi-talker speech input, that presents another mismatch when presented with single-talker speech input. In this paper, we propose an end-to-end neural network architecture for target speaker verification (tSV), which can effectively overcome the aforementioned mismatches.

Specifically, the proposed tSV system consists of three main components: a speaker attention module, a speaker representation module, and a speaker verification decision module. The speaker attention module extracts the target speaker's voice, that is further encoded by the speaker representation module into a discriminative speaker embedding for effective speaker verification. Following the multi-task learning methodology, we propose to jointly optimize the speaker attention module and the speaker representation module by simultaneously minimizing a signal reconstruction loss and a speaker identity loss. For the speaker verification decision module, a PLDA~\cite{prince2007probabilistic} classifier is trained with the speaker embeddings derived from the speaker representation module. % from the training set. %instead of x-vector. 
%At run-time, the speaker embeddings of both the enrollment utterance and the test utterance, either single- or multi-talker speech, are obtained from the same speaker representation module. Finally, the PLDA classifier makes the speaker verification decision by comparing these two speaker embeddings. %of the target enrollment utterance and the test utterance. 

%, because the target speaker extraction is only trained with overlapped multi-talker speech without knowing the single speaker's voice. With the superior performance of neural network based SV systems, such as, x-vector \cite{snyder2016deep}, d-vector \cite{wan2018generalized} and other similar feature representations \cite{huang2018angular},
%, i.e., SpEx+~\cite{ge2020spex+}, and a \textcolor{red}{speaker embedding back-end}. The speaker embedding back-end takes either the extracted speaker representation or the reconstructed speech sample of the target speaker as input, and generates speaker embedding for speaker verification decision.

%Meanwhile, this system could solve the SV problem in both overlapped multi-talker and single speaker conditions. The system becomes a universal single and multi-talker SV (tSV) system.

%Various speaker embedding back-ends are investigated in either time-domain or frequency-domain, which results in four different approaches. The front-end and back-end modules are jointly optimized with a scale-invariant signal-to-distortion ratio (SI-SDR) loss and cross-entropy loss via a multi-task learning framework. 

In this work, we seek to develop a unified framework for both single- and multi-talker speech inputs, that is fulfilled by exposing the tSV system to both single- and multi-talker input speech during training. This marks a departure from the single-talker assumption in traditional speaker verification systems, such as x-vector~\cite{snyder2016deep}, d-vector~\cite{wan2018generalized}, and other speaker representation techniques~\cite{huang2018angular}. In summary, this paper makes the following contributions:
\begin{enumerate}
\item We propose a unified network architecture that performs target speaker verification (tSV) for both single- and multi-talker speech. 
%\item We propose two architectures which result in four approaches with various time-domain and frequency-domain back-end speaker representation modules.
\item We propose a multi-task learning algorithm for the proposed tSV framework, which jointly optimize the target speaker attention module and the speaker representation module. 
\item We perform comprehensive studies on each individual network modules as well as at the system level. We successfully show that the proposed tSV framework not only works well under the multi-talker condition, but also performs as competitively as traditional SV under the single-talker condition.  %different corpora together with a cross corpus evaluation.
\end{enumerate}

The remainder of the paper is organized as follows. In Section~\ref{sec:system}, we introduce the proposed target speaker verification (tSV) system in details. In Section~\ref{sec:exp}, we describe the experiments through which we systematically evaluate the proposed tSV framework. In Section~\ref{sec:exp_res}, we report the experimental results. Finally, we conclude the study in Section~\ref{sec:con}.

% \vspace{-5pt}
\section{Unified Target Speaker Verification System for Single and Multi-talker Speech}
\label{sec:system}
% \vspace{-5pt}

\begin{figure}[tb]
\centering
\includegraphics[width=\linewidth]{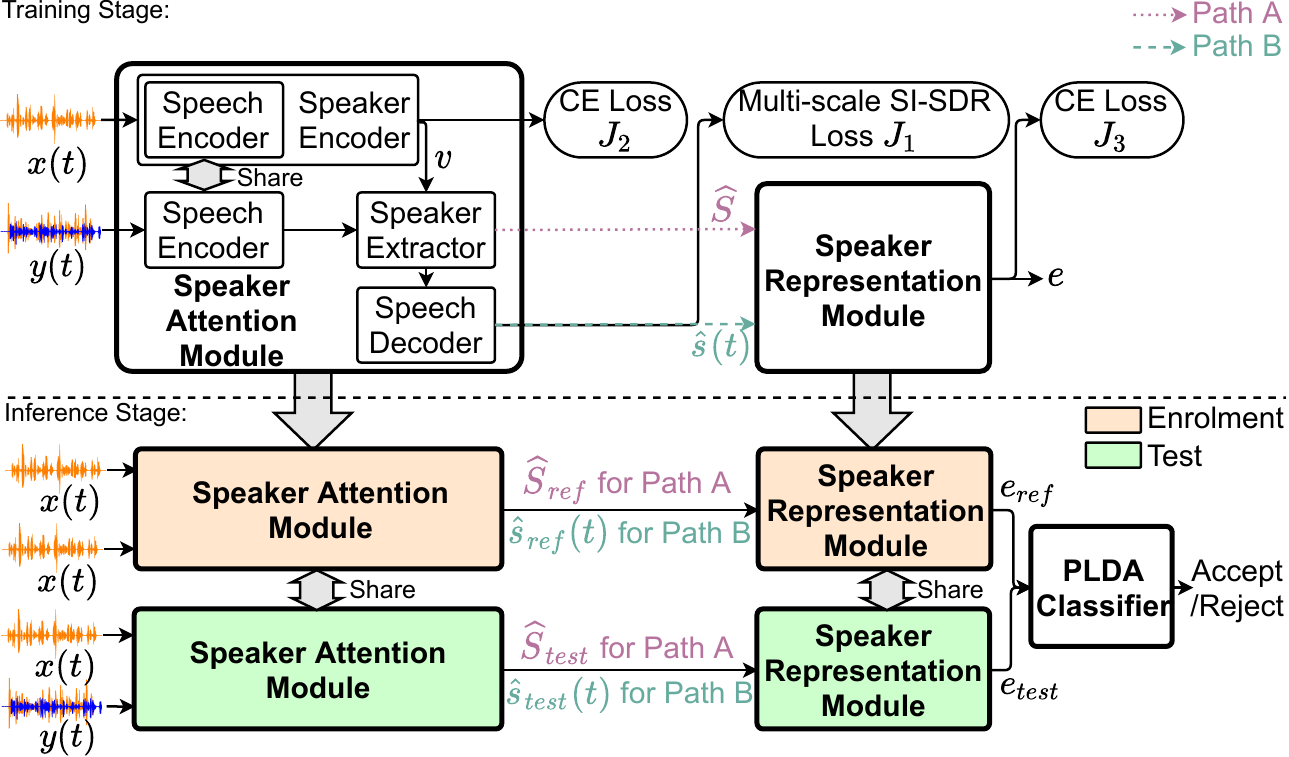}
\caption{The proposed tSV system consists of a speaker attention module, a speaker representation module, and a speaker verification decision module (PLDA classifier). $x(t)$ and $y(t)$ denote the single-talker reference speech and the observed speech during the training stage, whereas they are also used to denote the enrollment utterance and test utterance during the inference stage. \textit{Path A} and \textit{Path B} represent two output paths of the speaker attention module.}
%the enrolled and test speeches, respectively.}
\vspace{-10pt}
\label{fig:system_sesv}
\end{figure}

%\subsection{Overall System Architecture}
%\label{ssec:tSV}
As shown in Figure \ref{fig:system_sesv}, the proposed tSV system consists of a speaker attention  module, a speaker representation module, and a PLDA classifier. During training, let $x(t)$ be a reference speech from the target single talker, and $y(t)$ be the observed speech that could come from either a single talker or multiple talkers. The speaker attention module is trained to extract the target speech from $y(t)$ with the reference of $x(t)$. The extracted target speech is taken by the speaker representation module either in the form of an internal representation $\hat{S}$ (\textit{Path A} in Figure \ref{fig:system_sesv}) or as a reconstructed speech $\hat{s}(t)$ (\textit{Path B} in Figure \ref{fig:system_sesv}), and further transformed into an appropriate speaker embedding $e$. We train the speaker attention module and the speaker representation module jointly via a multi-task learning algorithm that will be explained in details in Section \ref{ssec:training_scheme}. %We train the speaker representation module through a speaker classification task with a cross-entropy loss. 
Finally, a PLDA classifier is trained on the derived speaker embeddings $e$ to perform speaker verification. %\textcolor{red}{(I suggest that we change $\hat{x}(t)$ to $\hat{s}_{ref}(t)$ and $\hat{s}(t)$ to $\hat{s}_{test}(t)$, and  $e_{\hat{x}}$ to $e_{\hat{s}_{ref}}$ , $e_{\hat{s}}$ to $e_{\hat{s}_{test}}$ in Figure 1)} \textcolor{blue}{(I have changed as in the Figure 1, slightly different from your suggestion.)}

During inference, let  $x(t)$ be an enrollment utterance, a.k.a the reference speech used for speaker attention modeling, from the target single talker, and $y(t)$ be the test utterance for the target speaker verification. %that could come from either a single talker or multiple talkers %either with the presence of the target speaker or without.
We leverage the trained speaker attention module and speaker representation module to derive an enrolled speaker embedding $e_{ref}$ for $x(t)$, and a target speaker embedding $e_{test}$ for $y(t)$. It is commonly assumed that the enrollment utterance is pre-recorded from the target single speaker. Therefore, it is not necessary for the enrollment utterance to pass through the speaker attention module. However, to keep the input to the speaker representation module consistent across both training and inference stages, we pass both the enrollment utterance and the test utterance through the shared speaker attention module as illustrated in Figure \ref{fig:system_sesv}. This is referred to as the standard inference configuration.

Under this configuration, $x(t)$ serves as the input reference speech, and at the same time, the input observed speech to obtain an internal representation $\hat{s}_{ref}(t)$ or $\hat{S}_{ref}$ for the enrolled target speaker. Besides, $x(t)$ is also served as the reference speech to extract a target speaker representation $\hat{s}_{test}(t)$ or $\hat{S}_{test}$ for the test utterance $y(t)$. In the case where the target speaker is absent from $y(t)$, no speech content of the target speaker is expected in   $\hat{s}_{test}(t)$ or $\hat{S}_{test}$.

%Under this configuration, $x(t)$ serves as both the input reference speech, and at the same time, the input observed speech to the speaker attention module. This module is expected to obtain $\hat{s}_{ref}(t)$ or $\hat{S}_{ref}$ for the enrolled target speaker. Similarly, the speaker attention module is expected to extract the speech content of the target speaker $\hat{s}_{test}(t)$ or $\hat{S}_{test}$ from the test utterance $y(t)$. 

%The inference procedure in Figure \ref{fig:system_sesv} is referred to as the standard configuration.

The speaker representation module then encodes $\hat{s}_{ref}(t)$ and $\hat{s}_{test}(t)$ (\textit{Path B}), or $\hat{S}_{ref}$ and $\hat{S}_{test}$ (\textit{Path A}), into speaker embeddings $e_{ref}$ and $e_{test}$, respectively. $e_{ref}$ encodes the enrolled target speaker, and  $e_{test}$ encodes the extracted target speaker as long as he/she is present in $y(t)$.  Finally, the PLDA classifier compares $e_{ref}$ and $e_{test}$ to accept or reject the speaker identity claim.

\subsection{Speaker Attention Module}
\label{ssec:target_speaker_extraction}

\begin{figure}[tb]
\centering
\includegraphics[width=0.9\linewidth]{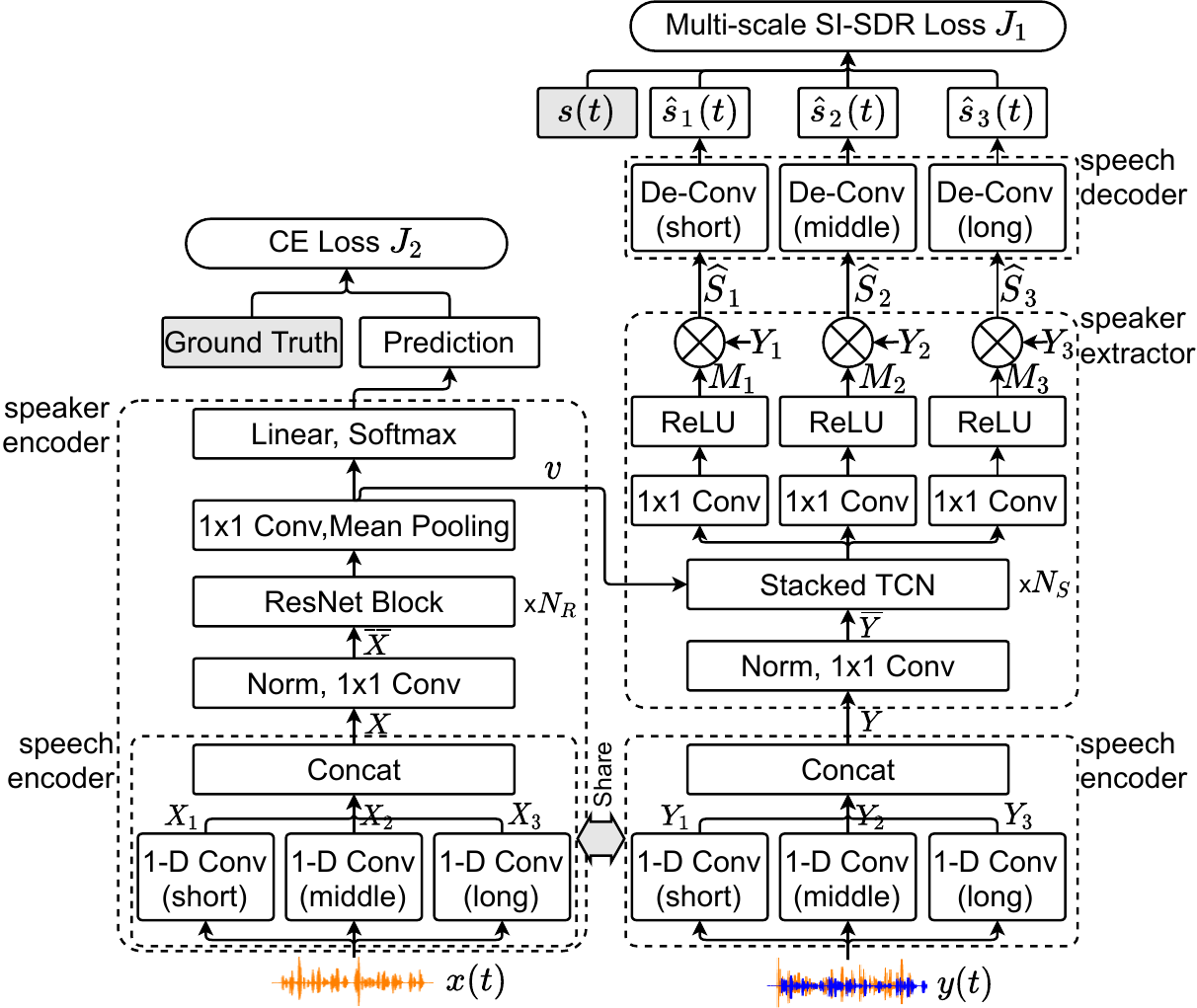}
\caption{Illustration of the speaker attention module that extracts the target speaker's voice from $y(t)$ with respect to the reference speech $x(t)$. The outputs are represented either in the form of an internal representation $\hat{S}$ or as a reconstructed speech $\hat{s}(t)$.}
\vspace{-10pt}
\label{fig:speaker_extraction}
\end{figure}

The speaker attention module in Figure \ref{fig:system_sesv}  consists of a shared \textit{speech encoder}, a \textit{speaker encoder}, a \textit{speaker extractor}, and a \textit{speech decoder}~\cite{xu2020spex}. As a more detailed illustration shown in Figure \ref{fig:speaker_extraction}, the \textit{speech encoder} encodes the observed speech $y(t)$, either single-talker or multi-talker, into spectrum-like embedding coefficients. Built on top of the shared \textit{speech encoder}, the \textit{speaker encoder} encodes the reference speech $x(t)$ into an utterance-level latent representation $v$ that represents the target speaker. %It guides the speaker extractor to pay selective attention only to the target speaker. %(we should move this to where we mention e. It is different from the speaker embedding $e$ in the back-end speaker representation module for SV, as shown in Figure \ref{fig:system_sesv}.
It guides the \textit{speaker extractor} to estimates an auditory mask for the target speaker, which only lets pass the target speaker's voice. Finally, the \textit{speech decoder} reconstructs the time-domain speech signal from the modulated embedding coefficients of the observed speech.

% \vspace{-5pt}
\subsubsection{Speech Encoder}
\label{sssec:encoder}
% \vspace{-5pt}
Inspired by the multi-scale time domain speech encoder introduced in \cite{xu2020spex}, we use three parallel 1-D convolutional layers, each has a different temporal resolution, to encode the speech inputs $y(t) \in \mathbb{R}^{1 \times T_1}$ and $x(t)\in \mathbb{R}^{1 \times T_2}$. Each of these convolutional layers has $N$ filters but with different kernel sizes of $L_1 (short),L_2(middle),L_3(long)$, respectively. The rectified linear unit (ReLU) activation function is used in these layers to produce non-negative embedding coefficients as the following %$Y=[Y_1, Y_2, Y_3]$ and $X=[X_1, X_2, X_3]$. %by transforming the time-domain speech sample, either a mixture speech $y(t) \in \mathbb{R}^{1\times T_1}$ or a single talker's voice $x(t)\in \mathbb{R}^{1\times T_2}$, at three different temporal resolution.
\begin{equation}
    Y_{i} = \text{ReLU}(y \ast U_i), \quad i=1,2,3
\end{equation}
\begin{equation}
    X_{i} = \text{ReLU}(x \ast U_i), \quad i=1,2,3
\end{equation}
where $\ast$ denotes the 1-D convolutional operator. $U_i\in \mathbb{R}^{N\times L_i}$ refers to the $N$ convolutional filters that has a kernel size of $L_i$ each. These filters are applied to the speech inputs at a fixed stride of $L_1/2$ samples to allow easy concatenation of the filter outputs at different temporal resolutions. $Y_i \in \mathbb{R}^{N\times K_1}$ and $X_i \in \mathbb{R}^{N\times K_2}$ are the encoded embedding coefficients, wherein $K_1=2(T_1-L_1)/L_1+1$ and $K_2=2(T_2-L_1)/L_1+1$ are the number of output frames. The outputs from these three parallel convolution layers are aligned and concatenated into multi-scale embedding coefficients $Y \in \mathbb{R}^{3N\times K_1}$ or $X\in \mathbb{R}^{3N\times K_2}$.
%with a window size of $L_i$ shifting every $L_1/2$ samples. $Y_i \in \mathbb{R}^{N\times K_1}$ and $X_i \in \mathbb{R}^{N\times K_2}$ are the encoded embedding coefficients.  $U_i\in \mathbb{R}^{N\times L_i}$ is the shared encoder basis. The multi-scale features representations $Y \in \mathbb{R}^{3N\times K_1}$ or $X\in \mathbb{R}^{3N\times K_2}$ are aligned and concatenated by keeping the same stride, $L_1/2$, across different scales.

\subsubsection{Speaker Encoder}
\label{sssec:embedding_network}
\begin{figure}[tb]
\centering
\includegraphics[width=0.8\linewidth]{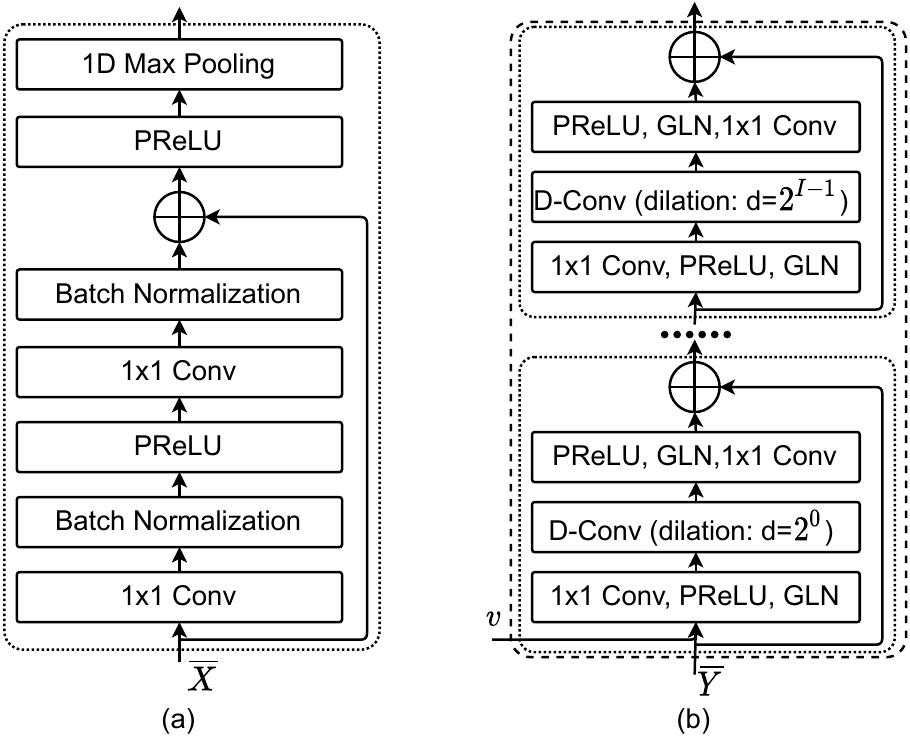}
\caption{(a) The diagram of a ResNet block. ``$1\times 1$ Conv" is a 1-dimensional convolution layer with the kernel size of $1\times 1$. ``$\bar{X}$" is the normalized embedding coefficients of the reference speech $x(t)$. (b) The diagram of a stacked TCN module that has $I$ blocks, each of which has a dilated depthwise convolution layer ``D-Conv"  with dilation ratio of $2^{i-1}$, $i=1,...I$. ``GLN" refers to the global layer normalization.  The target speaker representation $v$ is only presented at the first block. ``$\bar{Y}$" is the normalized embedding coefficients of the observed speech $y(t)$.}
\vspace{-10pt}
\label{fig:tcn}
\end{figure}

Following SpEx+ \cite{ge2020spex+}, we introduce a speaker encoder to encodes the reference speech $x(t)$ into an utterance-level representation $v$, which can characterize the voiceprint of the target speaker. %As $v$ is used to guide the extraction of the target speaker from the observed speech $y(t)$, the same speech encoder is used to encode the observed speech and the reference speech into multi-scale embedding coefficients $Y$ and $X$ by sharing the same feature space.
Built on top of the shared speech encoder, a 1-D convolutional layer with a kernel size of $1\times 1$, that is called a $1\times 1$ Conv, is applied to the normalized (channel-wise) embedding coefficients, followed by $N_R$ identical residual network (ResNet) blocks to progressively modify the representation. The details of each residual block are provided in Figure \ref{fig:tcn}(a). %As shown in Figure \ref{fig:tcn} (a), each ResNet block consists of two $1\times 1$ Conv layers and one 1-D max-pooling layer that has a kernel size of $1\times 3$. 
Finally, a $1\times 1$ Conv layer together with a mean pooling is used to transform the frame-level feature representations into a fixed dimensional utterance-level representation $v \in \mathbb{R}^{D \times 1}$. To summarise, the speaker encoder can be expressed as a function $g(\cdot)$ of the input speech $x(t)$,
\begin{equation} \label{Eq:latent-rep}
    v=g(x(t))
\end{equation}

To train this speaker encoder, we attach an output layer to classify the speaker identity of each utterance, which is trained jointly with other modules within a multi-task learning framework as will be explained in Section \ref{ssec:training_scheme}. It worth mentioning that the output layer and the loss function are not required during inference. 
%During training, an output layer is further attached to compute the cross-entropy loss between the ground truth and predicted speaker labels of each utterance. The cross-entropy loss forms part of multi-task learning at system level. The training of the speaker encoder will be further discussed in Section \ref{ssec:training_scheme}. At run-time inference, the output layer and loss calculation are not involved.

\subsubsection{Speaker Extractor}
\label{sssec:extractor}

The speaker extractor estimates a selective filter (i.e., mask), that only lets pass speech content related to the target speaker while masks off that of other interference  speakers. To this end, the multi-scale embedding coefficients $Y$ of the observed speech is first normalized channel-wise before being filtered by a $1\times 1$ Conv layer with $O$ filters. Then, stacked temporal convolutional network (TCN) modules are repeated $N_S$ times to transform the input embedding into an effective representation for mask estimation. The details of one such stacked TCN module are provided in Figure \ref{fig:tcn} (b), wherein $I$ dilated depthwise separable convolution layers are stacked with dilation factors of $2^{(i-1)}, i=1,2,...,I$. The depthwise convolution with $P$ filters and a kernel size of $1 \times Q$ is coupled with a $1\times 1$ Conv layer with $O$ fiters and global layer normalization (``GLN") to keep the total number of parameters manageable. 

%Then $I$ temporal convolutional network (TCN) are stacked into a block, as shown in Figure \ref{fig:tcn} (b). Each TCN applies a dilated depthwise separable convolution with $P$ filters and a kernel size of $1 \times Q$ to keep a manageable number of parameters. The dilation factor for each TCN is $2^{(i-1)}, i=1,2,...,I$. 

To achieve target speaker extraction, the mask estimation is conditioned on the target speaker representation $v$ derived from the speaker encoder during both training and inference. Specifically, the representation $v$ is repeated and concatenated to the normalized embedding coefficients $\bar{Y}$ as the inputs $\hat{Y} \in \mathbb{R}^{(O+D)\times K_1}$ to each stacked TCN module. Next, three parallel $1 \times 1$ convolutions layers with $N$ filters each are used to estimate mask $M_i\in \mathbb{R}^{N\times K_1}, i=1,2,3$, one for each encoded embedding coefficient $Y_i\in \mathbb{R}^{N\times K_1}$.

%To achieve the target speaker extraction, the mask estimation is conditioned on the target speaker representation $v$ derived from the speaker encoder during both training and inference. The representation $v$ is concatenated repeatedly to the normalized embedding coefficients. The embedding coefficients with repeated $v$ are used as the inputs $\hat{Y} \in \mathbb{R}^{(O+D)\times K_1}$ of the first TCN in each stacked TCN block. The stacked TCN block is repeated for $N_S$ times to learn a good embedding coefficient $\Tilde{Y} \in \mathbb{R}^{O\times K_1}$ for the following mask estimation. Three parallel $1\times 1$ convolutions with $N$ filters are then used to estimate a mask $M_i\in \mathbb{R}^{N\times K_1}, i=1,2,3$, one for each encoded embedding coefficient $Y_i\in \mathbb{R}^{N\times K_1}, i=1,2,3$ through a ReLU activation function.

Finally, the modulated embedding coefficients $\hat{S}_i \in \mathbb{R}^{N\times K_1}$ of the target speaker are obtained, for different scale $i=1,2,3$, by applying the estimated mask $M_i \in \mathbb{R}^{N\times K_1}$ onto the corresponding embedding coefficients $Y_i \in \mathbb{R}^{N\times K_1}$, 
\begin{equation} \label{Eq:Si}
\begin{aligned}
    \hat{S}_i &= M_i \otimes Y_i \\
    &= f(Y, v) \otimes Y_i
\end{aligned}
\end{equation}
where $\otimes$ denotes the element-wise multiplication, and $f(\cdot)$ represents the operations of the speaker extractor. %$Y$ is the multi-scale embedding coefficients.
\subsubsection{Speech Decoder}
\label{sssec:decoder}

The speech decoder reconstructs the time-domain signal $\hat{s}_i \in \mathbb{R}^{1\times T_1}$ from the modulated embedding coefficients $\hat{S}_i\in \mathbb{R}^{N\times K_1}$ through a de-convolutional layer that performs 1-D transposed convolution operations as follows,
\begin{equation}
    \hat{s}_i = \hat{S}_i \ast V_i, \quad i=1,2,3
\end{equation}
where $V_i\in \mathbb{R}^{N\times L_i}$ is the decoder basis.

\subsection{Speaker Representation Module}

Speaker representation module seeks to encode the speaker characteristics of an utterance of variable duration into a fixed-length vector, that is called speaker embedding, for example, x-vector~\cite{snyder2016deep}. The speaker representation with a fixed-length vector greatly facilitates the speaker comparison. In this paper, the speaker embedding, denoted as $e$, is derived from a speaker attention-representation pipeline, therefore, it is called target-speaker-vector, or ts-vector for short. It should be noted that the speaker embedding $e$, as the output of the speaker representation module, is derived for speaker comparison. Here, $e$ is not to be confused with the reference utterance latent representation $v$ in Eq. \ref{Eq:latent-rep}, that is derived for the speaker attention purpose. 

As introduced in the earlier section, the speaker attention module is trained to produce time-domain speech signal for the target speaker. As reported in~\cite{xu2020spex}, for speaker attention module with multi-scale speech encoding, the quality of the reconstructed speech signal $\hat{s}_1(t)$ with a high temporal resolution (a small kernel size) outperforms $\hat{s}_2(t)$ and $\hat{s}_3(t)$ with middle and low temporal resolutions, as shown in Figure \ref{fig:speaker_extraction}. 
Therefore, $\hat{s}_1(t)$ from the speaker attention module is chosen in this paper as the time-domain input to the speaker representation module (\textit{Path B} in Figure 1). Meanwhile, we also study the use of modulated embedding coefficients $\hat{S}$ (\textit{Path A} in Figure 1) as the input to the speaker representation module. We next discuss four different ts-vector speaker embedding schemes derived from them.

%the speaker attention module also produces  modulated embedding coefficients~\cite{xu2020spex}, denoted as $\hat{S}$ in Figure 1 (\textit{Path A} in Figure 1).  

\begin{figure*}[tb]
\centering
\includegraphics[width=\linewidth]{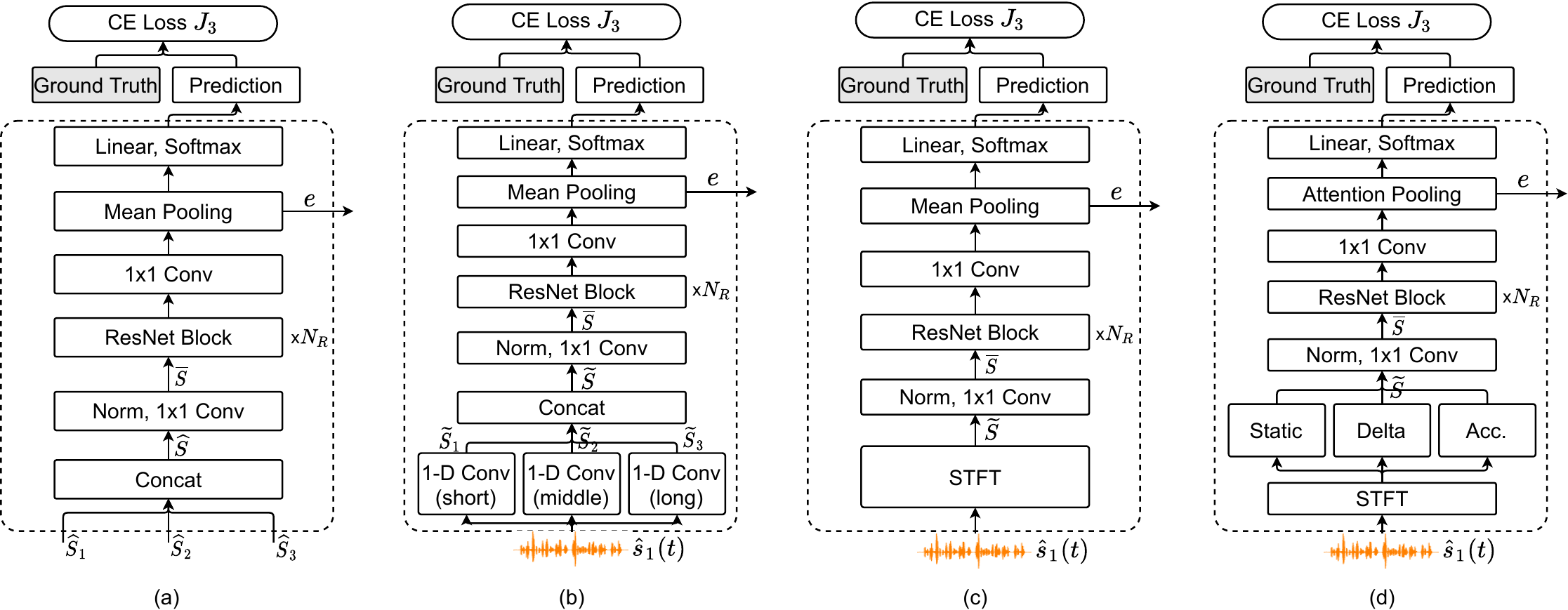}
\caption{The implementation of the speaker representation module with four speaker embedding schemes, where tSV-R follows \textit{Path A} of tSV system, while tSV-T, tSV-F, and tSV-FA follow \textit{Path B} of tSV system. (a) tSV-R with ts-vector-R; (b) tSV-T with ts-vector-T; (c) tSV-F with ts-vector-F; (d) tSV-FA with ts-vector-FA.}
\vspace{-10pt}
\label{fig:speaker_embedding}
\end{figure*}

\subsubsection{Speaker Embedding with Embedding Coefficients $\hat{S}$}
\label{sssec:extractor1}
As shown in Figure \ref{fig:speaker_embedding} (a), following a similar network architecture as the speaker encoder introduced in Section \ref{sssec:embedding_network}, we design a network to transform the temporal sequence $\hat{S}$ into a speaker embedding. 
To ensure a fixed dimensional output from the network, we use a mean pooling layer to read out the target speaker embedding $e$, or ts-vector. We refer to the resulting speaker embedding $e$ as ts-vector-R, and the tSV system as tSV-R hereafter.
%This network transformation can be summarised by the following equation,
%\begin{equation}
%    e = h_1(\hat{S})
%\end{equation}

%The speech data from \textit{Path A} is a sequence of modulated embedding coefficients $\hat{S}$, that represent the target single-talker's speech, and resemble the spectrogram. To extract a speaker embedding, a network structure, $h_1(\cdot)$, is required to project the temporal sequence $\hat{S}$ into a fixed dimensional speaker embedding $e$. The detail is illustrated in Figure \ref{fig:speaker_embedding} (a),

%The network $h_1(\cdot)$ adopts a similar architecture as the speaker encoder in Section \ref{sssec:embedding_network}. As input embedding coefficients $\hat{S}$ are similar to the embedding coefficients $X$ in Figure \ref{fig:speaker_extraction}, which is already in an embedding space, a speech encoder is not required. We obtain the target speaker embedding $e$, or ts-vector, through the mean pooling layer. %The speaker representation module is optimized by a speaker classification task, that forms part of the multi-task learning at system level, as will be introduced in Section \ref{ssec:training_scheme}. 

\subsubsection{Speaker Embedding with Time-domain Encoder}
\label{sssec:extractor2}

%From the conclusions in \cite{xu2020spex}, the quality of the reconstructed signal $\hat{s}_1(t)$ with a high temporal resolution (a short window size in speech encoder) is better than the signals ($\hat{s}_2(t)$ and $\hat{s}_3(t)$) reconstructed from the middle and low temporal resolutions (middle and long window size), as shown in Figure \ref{fig:speaker_extraction}. Therefore, we expect better performance of the system that takes the best reconstructed signal $\hat{s}_1(t)$ than the system with extracted embedding coefficients $\hat{S}=[\hat{S}_1; \hat{S}_2; \hat{S}_3]$. 

We now take the best reconstructed signal $\hat{s}_1(t)$ from \textit{Path B} of the speaker attention module as the input to the speaker representation module. As the reconstructed speech $\hat{s}_1(t)$ is a time-domain signal, the time-domain processing is required to encode the speech signal. For fair comparison with $\hat{S}$ from \textit{Path A} that spans across multiple time scales, $\hat{s}_1(t)$ is first encoded by a time-domain encoder with the same temporal resolutions to $\tilde{S}$. For simplicity, we adopt the same speaker encoder as introduced in Section \ref{sssec:embedding_network}, which consists of a trainable speech encoder front-end.
%the time-domain speaker representation architecture $h_2(\cdot)$, which is same as the speaker encoder as introduced in Section \ref{sssec:embedding_network}. 
%The input $x(t)$ in Figure \ref{fig:speaker_extraction} is replaced as the best reconstructed signal $\hat{s}_1(t)$,
As illustrated in Figure \ref{fig:speaker_embedding} (b), the encoded multi-scale representation is further transformed into a speaker embedding using the same network structure as adopted for tSV-R. We refer to the resulting speaker embedding $e$ as ts-vector-T, and the tSV system as tSV-T hereafter.

%begin{equation}
%    e = h_2(\hat{s}_1(t))
%\end{equation}

\subsubsection{Speaker Embedding with Frequency-domain Encoder}
\label{sssec:extractor3}
We further use $\hat{s}_1(t)$ from \textit{Path B} of the speaker attention module as the input but in a different way. As illustrated in Figure \ref{fig:speaker_embedding} (c), instead of using a trainable time-domain speech encoder, we study the use of frequency domain processing via the short-time Fourier transform (STFT) analysis.
%When the best reconstructed signal $\hat{s}_1(t)$ is used as the input for the following back-end speaker representation module, the time-domain signal $\hat{s}_1(t)$ could be also encoded into frequency-domain spectral features via a STFT, instead of the trainable speech encoder used in Section \ref{sssec:extractor2}.
%Unlike the multi-scale the speech encoder $h_2(\cdot)$, the frequency-domain processing only adopts a single scale magnitude spectrum. 
Comparing to the tSV-T shown in Figure \ref{fig:speaker_embedding} (b), we note that, besides the difference in speech encoder front-end, the frequency-domain speaker representation module shares the same architecture with that of tSV-T. The resulting speaker embedding $e$ is referred to as ts-vector-F, and the tSV system as tSV-F hereafter.

%The speaker representation module is optimized by a speaker classification task, that forms part of the multi-task learning at system level.

\subsubsection{Speaker Embedding with Frequency-domain Attention}
\label{sssec:extractor4}
For frequency-domain SV, i.e., i-vector PLDA \cite{Dehak&Kenny2011, Kenny10, Garcia11}, the static and dynamic features are shown to improve the effectiveness of speaker embeddings, for example, mel-frequency cepstral coefficients (MFCC) with its delta and acceleration \cite{Rao&Mak2013a}. In addition, when deriving an utterance-level speaker embedding from a frame-level representation, studies have shown the superiority of using the attention mechanism to replace the mean pooling~\cite{okabe2018attentive,wang2018attention}. Therefore, we extend the frequency-domain speaker representation module tSV-F by 1) including delta and acceleration of the magnitude spectrum as additional features, and 2) replacing the mean pooling layer with an attentive statistic pooling layer. The details of the proposed model are illustrated in Figure \ref{fig:speaker_embedding} (d). We refer to the resulting speaker embedding $e$ as ts-vector-FA, and the tSV system as tSV-FA hereafter.

To train these speaker representation modules, we leverage a speaker classification task that forms part of the multi-task learning framework as introduced in the following section. 

\subsection{Multi-task Learning}
\label{ssec:training_scheme}

Speaker attention module primarily seeks to enhance the perceptual quality of the target speaker, which does not directly contribute to the improvement of speaker discrimination at the speaker representation module. Therefore, to unify their contributions, we propose to jointly optimize the speaker attention module and the speaker representation module via a multi-task learning framework. In the following, we will explain how multiple loss functions are contributed individually as well as at the system level to improve the speaker discrimination. 

%, by simultaneously minimizing a signal reconstruction loss and a speaker discrimination loss.
As discussed in Section \ref{ssec:target_speaker_extraction}, the main objective of the speaker attention module is to extract target speaker's voice with a high perceptual quality. This can be achieved by optimizing a multi-scale scale-invariant signal-to-distortion ratio (SI-SDR) loss \cite{le2019sdr}, denoted as $J_1$.
\begin{equation}\label{eq:J1}
    J_1 = -[(1-\alpha-\beta) \rho(\hat{s}_1, s)+\alpha \rho(\hat{s}_2,s)+\beta \rho(\hat{s}_3,s)]
\end{equation}
\noindent{where $\alpha$ and $\beta$ are tunable hyperparameters to adjust the contributions at different temporal scales. $\hat{s}_1$, $\hat{s}_2$ and $\hat{s}_3$ are the reconstructed signals from modulated embedding coefficients $\hat{S}_1$, $\hat{S}_2$ and $\hat{S}_3$, respectively. $s$ is the target clean signal. The SI-SDR loss \cite{le2019sdr}, denoted as $\rho(\cdot, \cdot)$, measures the error between the reconstructed and the target clean signals.}
\begin{equation}\label{eq:sisdr}
    {\rho}(\hat{s}, s) = 10\log_{10}\left(\frac{||\frac{\langle\hat{s}, s\rangle}{\langle s,s\rangle}s||^2}{||\frac{\langle\hat{s}, s\rangle}{\langle s,s\rangle}s-\hat{s}||^2}\right)
\end{equation}
where $\langle\cdot,\cdot\rangle$ indicates the inner product. In order to ensure scale invariance, the signals $\hat{s}$ and $s$ are normalized to zero-mean in the above formulation. 

Since speaker representation $v$ is crucial to the success of speaker extraction, we further introduce a speaker classification task~\cite{xu2020spex} to the speaker encoder within the speaker attention module, so as to improve the quality of speaker representation $v$. Specifically, the cross-entropy loss $J_2$ is applied to the speaker classification task.

%As we would also like the speaker extraction to improve the ultimate speaker verification, we further introduce a cross-entropy loss for speaker classification~\cite{xu2020spex} to the speaker attention module, denoted as $J_2$. They are illustrated in Figures \ref{fig:system_sesv} and \ref{fig:speaker_extraction}. 
\begin{equation} \label{eq:J2}
    J_2 = -\sum_{c=1}^{C} p_{c}\log(P(c|x))
\end{equation}
where $C$ is the total number of speakers in the speaker classification task. $p_c$ is 1 if the segment $x$ belongs to speaker $c$, otherwise, $p_c$ is 0. $P(c|x)$ is the probability for the segment $x$ to be speaker $c$. Besides optimized for the speaker classification task through $J_2$, the speaker encoder is also jointly optimized for speaker extraction via $J_1$.

%With the cross-entropy loss $J_2$, the speaker encoder is optimized to predict the correct speaker labels. It generates an utterance-level representation $v$ that is required by the speaker extractor. The speaker encoder is optimized to produce correct speaker label prediction, at the same time, extract the target speaker's voice with high SI-SDR together with other network components in the speaker attention module. %The workflow is illustrated in Figure \ref{fig:speaker_extraction}. 

The speaker representation module seeks to encode the enrollment utterance and the test utterance into speaker embeddings $e_{ref}$ and $e_{test}$ that are suitable for speaker comparison. To effectively characterize the speakers, the speaker representation module is trained together with the speaker attention module under another speaker classification task with a cross-entropy loss $J_3$, as shown in Figure \ref{fig:speaker_embedding},
\begin{equation} \label{eq:J3_S}
    J_3 = -\sum_{c=1}^{C} p_{c}\log(P(c|\hat{S}))
\end{equation}
or
\begin{equation} \label{eq:J3_s}
    J_3 = -\sum_{c=1}^{C} p_{c}\log(P(c|\hat{s}_1))
\end{equation}
where $\hat{S}=[\hat{S}_1;\hat{S}_2;\hat{S}_3]$ and $\hat{s}_1$ are the  modulated embedding coefficients and the reconstructed signal from the speaker attention module for \textit{Path A} and \textit{Path B}, respectively. 

%It takes the output of speaker attention module, either as the modulated embedding coefficients $\hat{S}=[\hat{S}_1;\hat{S}_2;\hat{S}_3]$ or the reconstructed signal $\hat{s}_1$, as shown in Figure 1, and detailed in Figure \ref{fig:speaker_embedding}.
%where $C$ is the number of speakers (classes) in the speaker classification task. $p_c$ is 1 if the segment $x$ is speaker $c$, otherwise, $p_c$ is 0. $P(c|\hat{S})$ is the probability when the extracted segment is predicted as speaker $c$ given the modulated embedding coefficients $\hat{S}$, as shown in Figure \ref{fig:speaker_embedding} (a). $P(c|\hat{s}_1)$ is the probability when the extracted segment is predicted as speaker $c$ given the reconstructed signal $\hat{s}_1$, as shown in Figure \ref{fig:speaker_embedding} (b), \ref{fig:speaker_embedding} (c) and \ref{fig:speaker_embedding} (d).

Finally, we jointly optimize the proposed framework with a total loss $J$, that is the weighted sum of $J_1$, $J_2$ and  $J_3$,
\begin{equation} \label{eq:totoal_loss}
    J = J_1 + \gamma J_2 + \eta J_3
\end{equation}
where $\gamma$ and $\eta$ are tunable hyperparameters to align different objectives.

\subsection{Target Speaker Verification}

The PLDA classifier~\cite{prince2007probabilistic} has been applied widely to many speaker embeddings, such as, i-vector \cite{Dehak&Kenny2011, Kenny10, Garcia11}, x-vector \cite{snyder2016deep,snyder2018x,snyder2019speaker}, and d-vector \cite{wan2018generalized}. With the four target speaker embedding schemes implemented in the speaker representation module, we are now ready to train a PLDA classifier for target speaker verification. 

% As ts-vectors share similar properties with other speaker embedding techniques, the PLDA classifier can thus benefit from the existing back-end technologies, such as length-normalization, PLDA scoring, and domain adaptation techniques. 
Specifically, prior to the PLDA classification, the speaker embeddings (ts-vectors) are centered following a linear discriminate analysis (LDA) for dimensionality reduction and a length normalization. To improve the system generalizability, we train the PLDA classifier using the extracted speaker embeddings from the training set that consists of both single talker and multi-talker speech samples. During the run-time inference, the PLDA classifier compares the enrolled target speaker and the extracted target speaker using their respective speaker embeddings, i.e.,  $e_{ref}$ and $e_{test}$. Finally, PLDA scores are
normalized with an adaptive s-norm \cite{sturim2005speaker}, and used to accept or reject the speaker identity claim.

\section{Experimental Setup}
\label{sec:exp}

%\vspace{-5pt}
\subsection{Speech Corpora}\label{sec:exp_data}
% \vspace{-2pt}

To evaluate the performance of the proposed target speaker verification system, we conducted experiments on two standard corpora\footnote{All the configurations (including the utterances list for training, development and evaluation) and the code could be found here:  \url{https://github.com/xuchenglin28/target_speaker_verification}}: 1) WSJ0~\cite{garofolo1993csr} and its 2-talker mixture version (WSJ0-2mix-extr)~\cite{xu2019optimization} with a total number of 119 speakers, and 2) LibriSpeech \cite{panayotov2015librispeech} and its 2-talker mixture version (Libri2Mix) \cite{cosentino2020librimix} with a total number of 1,212 speakers.

\subsubsection{WSJ0 and WSJ0-2mix-extr Corpora}
\label{sssec:wsj0}

WSJ0 corpus \cite{garofolo1993csr} consists of read speech from the Wall Street Journal. It has three original collections: ``si\_tr\_s'', ``si\_dt\_05'', and ``si\_et\_05''. In this work, we first generated a {single-talker dataset (WSJ0-1talker)} by randomly selecting speech from the WSJ0 corpus for single-talker SV. WSJ0-1talker consists of three subsets: training, development, and evaluation. Specifically, 11,560 utterances from a total of 101 speakers ($50$ male and $51$ female speakers) were selected from WSJ0 ``si\_tr\_s'' set, and split into the training set (8,769 utterances) and the development set (the rest 2,791 utterances). The evaluation set includes 1,857 enrollment utterances and 1,478 test utterances of 18 speakers ($10$ male and $8$ female speakers) from the WSJ0 ``si\_dt\_05'' and ``si\_et\_05'' sets. From this evaluation set, we further constructed $3,000$ target trials and $48,000$ non-target trials. The speech samples were down-sampled to 8kHz. 

WSJ0-2mix-extr \cite{xu2019optimization} corpus was generated from the WSJ0 Corpus by mixing two randomly selected utterances as a {2-talker dataset (WSJ0-2talker)}\footnote{The WSJ0-2mix-extr corpus simulation code is available at: \url{https://github.com/xuchenglin28/speaker_extraction}}. WSJ0-2talker also consists of training, development, and evaluation sets. Specifically, the training set included $20,000$ mixtures that were generated by mixing two randomly selected utterances from the aforementioned $101$ speakers in the WSJ0 corpus. Similarly, the development set with $5,000$ mixtures was also generated from the same $101$ speakers. The evaluation set has $3,000$ mixtures generated from the aforementioned $18$ different speakers as the test utterances, which are unseen during training. The signal-to-noise ratio (SNR) between the target speaker and the interference speaker of each mixture was randomly chosen between 0dB and 5dB. For the training and evaluation of the speaker attention module, each mixture has a corresponding reference speech and a single talker target speech. For the multi-talker SV evaluation, $3,000$ target trials and $48,000$ non-target trials were generated from the same $1,875$ enrollment utterances from WSJ0-1talker dataset and the above $3,000$ 2-talker test utterances.

\begin{figure}[t]
  \centering
  \centerline{\includegraphics[width=0.8\linewidth]{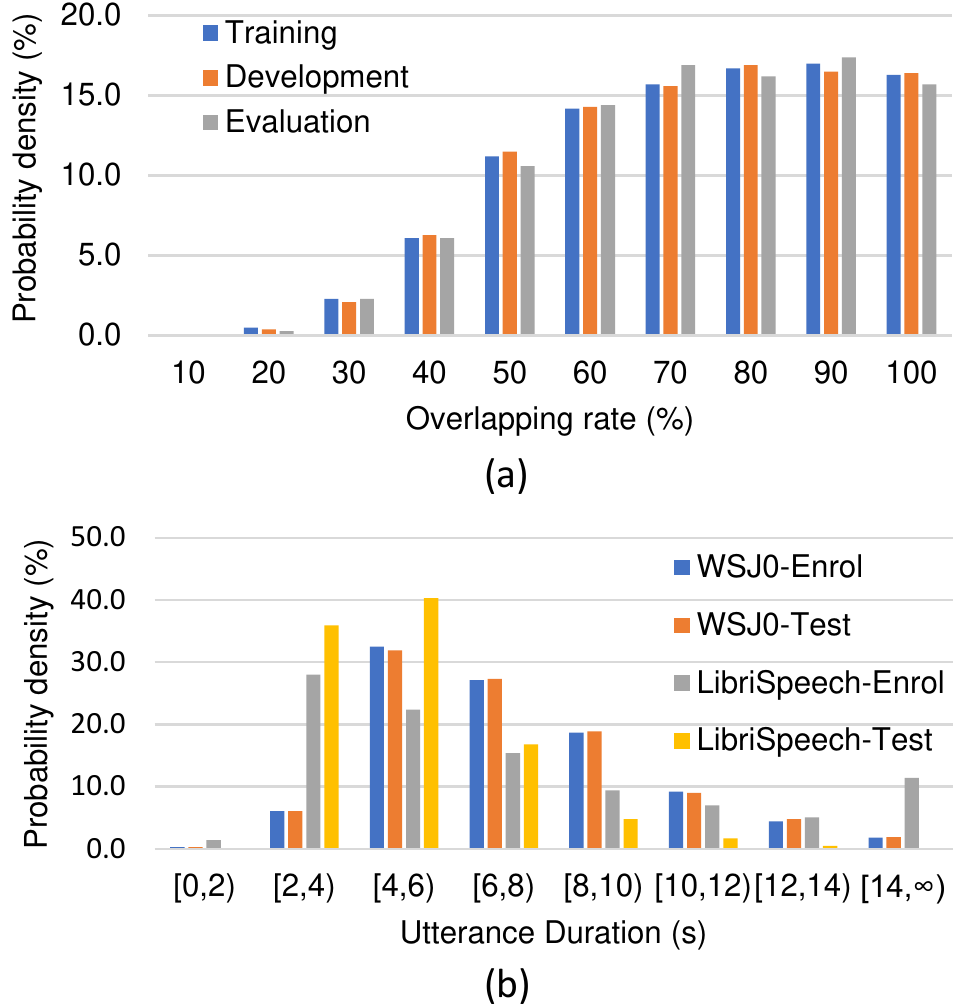}}
  %\vspace{-6pt}
    \caption{(a) The overlapping rate distribution of 2-talker mixture on training, development, and evaluation sets of the WSJ0-2mix-extr corpus. (b) The duration distribution of the evaluation set for both enrol and test utterances.  }
    \label{fig:overlap_ratio}
\vspace{-10pt}
\end{figure}

The mixtures in the WSJ0-2talker were generated according to the maximum duration protocol, where the shorter utterance was appended with zeros to match the duration of the longer utterance. The maximum duration protocol leads to various overlapping rates for the mixtures, as summarized in Figure~\ref{fig:overlap_ratio} (a). Most 2-talker speech samples are heavily overlapped. The average duration of enrollment and test utterances is $7.2$ second and $7.3$ second, respectively. The distribution of utterance duration for the enrollment and evaluation sets is summarized in Figure \ref{fig:overlap_ratio} (b). Most utterances used for both enrollment and test are less than 10.0 seconds.

In both the WSJ0-1talker and WSJ0-2talker, the speakers in the evaluation set are excluded from the training and development sets. 

\subsubsection{LibriSpeech and Libri2Mix Corpora}
\label{sssec:librispeech}

LibriSpeech \cite{panayotov2015librispeech} is derived from audiobooks that are part of the LibriVox\footnote{\url{https://librivox.org}} project. Libri2Mix \cite{cosentino2020librimix} is a clean 2-talker mixture corpus generated from the LibriSpeech corpus by mixing two randomly selected utterances. Libri2Mix\footnote{The simulation code of Libri2Mix is:\url{https://github.com/JorisCos/LibriMix}} consists of several subsets: ``train-360" with 50,800 utterances (212 hours, 921 speakers), ``train-100" with 13,900 utterances (58 hours, 251 speakers), ``dev" with 3,000 utterances (11 hours, 40 speakers), and ``test" with 3,000 utterances (11 hours, 40 speakers).

To study the speaker attention mechanism,  each 2-talker mixture speech could be reused in the following way. In the first instance, we take one speaker, e.g., \textit{Speaker A} as the target speaker, and another speaker e.g., \textit{Speaker B} as the interference speaker. By swapping the role of \textit{Speaker A} and \textit{Speaker B}, we can use the same speech sample in the second instance.

We generated a 2-talker dataset (Libri-2talker) with a training set, a development set, and an evaluation set. The training set (127,056 examples, 1,172 speakers) and development set (2,344 examples, 1,172 speakers) are randomly chosen from the ``train-360" and ``train-100" sets of the Libri2Mix corpus. The evaluation set includes 2,260 enrollment utterances from the ``test-clean" set of the LibriSpeech corpus and 6,000 test utterances from ``test" set of the Libri2Mix corpus. With these enrollment and test utterances, 6,000 target trials and 114,000 non-target trials were generated to evaluate the performance of proposed system for multi-talker SV.

We also generated a single-talker dataset (Libri-1talker), which consisted of training, development, and evaluation sets. The training set (125,925 utterances) and development set (6,628 utterances) were randomly selected from the original ``train-clean-100" and ``train-clean-360" of the LibriSpeech corpus with a total of 1,172 speakers and around 460 hours of clean speech. The evaluation set includes the same 2,260 enrollment utterances and 6,000 test utterances of target single-talker speech from the ``test" set in Libri2Mix corpus with 40 speakers. Similarly, 6,000 target trials and 114,000 non-target trials were generated to evaluate the performance of the SV system under the single speaker's condition.

The Libri2Mix was simulated according to the minimum duration protocol, where the longer utterance was cut short to match the duration of the shorter one. The minimum duration protocol leads to roughly $100\%$ overlapping rate. The average duration of enrollment and test utterances is $7.4$ second and $5.0$ second, respectively. The duration distribution of enrollment and test set is reported in Figure \ref{fig:overlap_ratio} (b). Most utterances for both enrollment and test are less than 8.0 seconds.

% \vspace{-5pt}
\subsection{Configuration of Speaker Attention Module}\label{sec:exp_tse}
% \vspace{-2pt}

We adopt the complete time-domain implementation~\cite{ge2020spex+} as the speaker attention mechanism for target speaker extraction. The shared speech encoder encodes both the mixture speech and the reference speech into multi-scale embedding coefficients by three parallel 1-D convolutions with N (=256) filters each. The three 1-D convolutions apply different filter lengths of $L_1$(=2.5ms), $L_2$(=10ms), $L_3$(=20ms) for complementary temporal resolutions. The multi-scale embedding coefficients are firstly normalized by their mean and variance with trainable gain and bias on the channel dimension in the speaker extractor. Then the $1\times 1$ convolution linearly transforms the normalized embedding coefficients to have $O$(=256) channels. Each TCN block has a dilated depthwise convolution with $P$ (=512) filters and a kernel size of $1\times Q$(=3). $I(=8)$ TCN blocks are stacked into a module and further repeated for $N_S(=4)$ times. The speech decoder reconstructs the modulated embedding coefficients into time-domain speech signals. The configuration of the de-convolution is kept the same as that used in the speech encoder. 

Since the extraction pipeline requires the information of the target speaker, the speaker encoder employs same normalization and $1\times 1$ convolution as in speaker extractor. Then the normalized embedding coefficients are used as inputs to the following $N_R$(=3) ResNet block. The ResNet block has two $1\times 1$ convolutions layers with $256$ filters each, and a 1-D max-pooling layer with the kernel size of $1\times 3$.

% \vspace{-5pt}
\subsection{Configuration of Speaker Representation and Speaker Verification Modules}\label{sec:exp_sv}
% \vspace{-2pt}

As shown in Figure \ref{fig:speaker_embedding}, the four speaker representation modules share a similar network architecture except for some variations in the speech encoding front-end. For tSV-R, it directly takes the modulated embedding coefficients $\hat{S}$ from \textit{Path A} as inputs. The tSV-T uses three parallel 1-D convolution blocks that has the same network configurations as the speech encoder in the speaker attention module. Both tSV-F and tSV-FA systems employ the STFT with a hamming window of size 32 ms (=256 samples), and a stride of 16 ms to perform signal analysis. In addition, the delta and acceleration features are calculated with an order of $2$ inside the tSV-FA system. 

These speech encoding front-ends are followed by channel-wise normalization, $1\times 1$ Conv with $256$ filters, and $N_R$(=3) stacked ResNet blocks. Each ResNet block has two $1\times 1$ Conv layers with $256$ filters and a 1-D max-pooling layer with the kernel size of $1\times 3$. The resulting output speaker embedding from the mean pooling layer has a dimension of 256. For tSV-FA, the attentive statistic pooling layer is formed by two fully-connected layers that have $500$ and $1$ neurons, respectively. The ReLU activation function is applied after the first layer. A softmax activation function is applied on the temporal dimension to obtain a normalized weight coefficient for each frame-level feature. The output speaker embedding, with a dimension of $512$, is generated by concatenating the weighted mean and standard deviation vectors.

For the speaker verification module, the LDA and Gaussian PLDA models with 100 latent variables are trained on the derived speaker embeddings from the training data. 

\subsection{Training Procedure}
We organized the training of the proposed tSV system into three stages. At the first stage, following the same configuration in \cite{ge2020spex+}, we trained the speaker attention module with 2-talker mixture speech segments that had a fixed duration of 4s. The module was trained with the weighted loss $J=J_1+\gamma J_2$, where $J_1$ was defined in Eq. \ref{eq:J1} and $J_2$ was defined in Eq. \ref{eq:J2}. The weights $\alpha$, $\beta$ in $J_1$ and $\gamma$ were fine-tuned to take the values of $0.1$, $0.1$, and $10$, respectively. The learning rate was initialized at $10^{-3}$, and halved whenever the loss was stagnant for 3 consecutive epochs. To ensure the front-end speaker attention module also works well under the single talk condition, we further fine-tuned the module with both the same 2-talker mixtures and the additional single speaker's speech at a reduced learning rate of $10^{-4}$. 

At the second stage, we froze the speaker attention module and further trained the speaker representation module using the extracted speech from both the 2-talker mixture and the single speaker. A learning rate of $10^{-4}$ was used at this stage. Lastly, to improve the synergy between the front-end speaker attention module and the back-end speaker embeddding extractor, they were jointly fine-tuned using the weighted loss $J=J_1+\gamma J_2+\eta J_3$ at a reduced learning rate of $10^{-5}$, where a same value of 10 were used for both $\gamma$ and $\eta$. The Adam algorithm \cite{kingma2014adam} was adopted to optimize the network across all three stages.

%\subsection{Contrastive enrollment Process for tSV System with Path B during Inference}
\subsection{Alternative Configuration for tSV Inference with Path B}

\begin{figure}[tb]
\centering
\includegraphics[width=\linewidth]{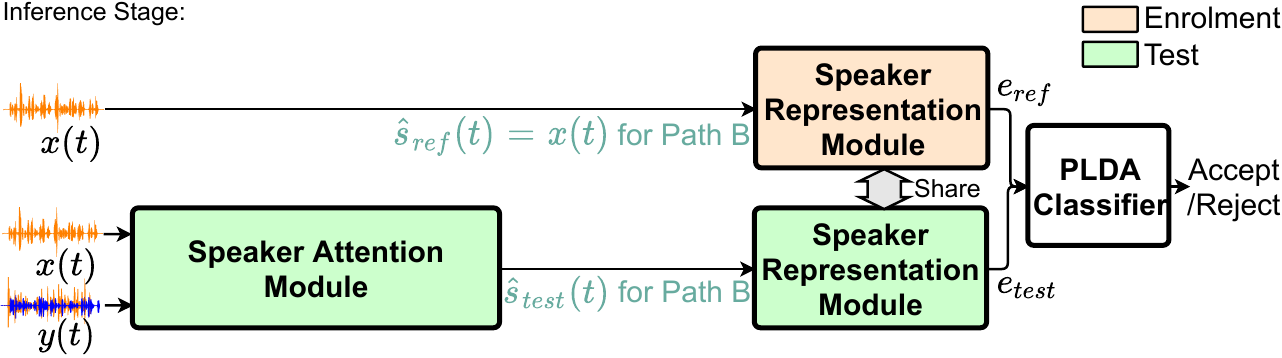}
\caption{An alternative inference flow-chart for the tSV system with \textit{Path B} in Figure \ref{fig:system_sesv}, where the enrollment utterance is directly taken by speaker representation module, bypassing the speaker attention module.}
\vspace{-10pt}
\label{fig:system_inference}
\end{figure}

In Figure \ref{fig:system_sesv}, to keep the same processing procedures for both $e_{ref}$ and $e_{test}$ during inference, we pass the enrollment utterance and the test utterance through the shared speaker attention module in the same way. We note an alternative implementation to \textit{Path B} is bypassing the speaker attention module and directly feeding the enrollment utterance into the speaker representation module. %It works for ts-vector-T, ts-vector-F, and ts-vector-FA speaker embeddings comparison,  as shown in Figure \ref{fig:system_inference}. 
 
 %As shown in Figure \ref{fig:system_sesv}, the speaker representation module in \textit{Path B} had either a trainable speech encoder for time-domain processing or a short-time Fourier transform (STFT) for frequency-domain processing. The tSV-T, tSV-F and tSV-FA systems could adopt the other inference flow-chart, as shown in Figure \ref{fig:system_inference}, where the enrollment utterance didn't go through the speaker attention module.
 
In this way, the speaker representation model can generate $e_{ref}$ directly from $x(t)$ instead of using its reconstructed version $\hat{s}_{ref}(t)$, as illustrated in Figure \ref{fig:system_inference}. The resulting speaker embedding $e_{ref}$ and $e_{test}$ can be compared by the PLDA classifier to make the speaker verification decision. However, $e_{ref}$ is derived from the enrollment utterance $x(t)$, while $e_{test}$ is derived from the extracted speech $\hat{s}_{test}(t)$. This marks a mismatch between their representations, which may lead to a sub-optimal decision. To evaluate the effect of this mismatch, we conducted an experiment on this alternative configuration, and reported the results in Section \ref{ssec:eval_enrol}.

\subsection{Evaluation Metrics}

The speaker attention module was evaluated using the SI-SDR \cite{le2019sdr} loss defined in Eq. \ref{eq:sisdr} for both 2-talker mixture and clean single speaker conditions. The speaker verification decision module (i.e., PLDA classifier) was evaluated using equal error-rate (EER), minimum of the normalized detection cost function (DCF) with P\_Target = $0.01$ (denoted as DCF08 \cite{martin2009nist}), and DCF with P\_Target = $0.001$ (denoted as DCF10 \cite{martin2010nist}), respectively.

\vspace{-2pt}
\section{Experimental Results} \label{sec:exp_res}
%\vspace{-2pt}

We start by validating the proposed speaker attention module and speaker representation module through experiments. We then evaluate the proposed tSV system with four target speaker embedding schemes on the
{ WSJ0 and WSJ0-2mix-extr corpora in Section \ref{ssec:eval_extraction} to Section \ref{ssec:eval_enrol}. We also compare the proposed tSV with other state-of-the-art solutions to multi-talker speaker verification on the WSJ0-2mix-extr corpus in Section \ref{ssec:eval_comp_other}. We further evaluate the proposed tSV system on LibriSpeech and Libri2Mix corpora, that are larger datasets, in Section \ref{ssec:eval_best_libri2mix}.  Finally, we evaluate our target speaker embedding scheme across WSJ0-2mix-extr and Libri2Mix corpora in Section \ref{ssec:eval_across}. }

%We firstly conduct experiments on WSJ0 and WSJ0-2mix-extr corpora, that are described in Section \ref{sssec:wsj0}. We compare among the four speaker embedding techniques, and against the competitive state-of-the-art methods. To evaluate the effectiveness of the tSV system on a large corpus, the best speaker embedding technique is further trained on LibriSpeech and Libri2Mix corpora with thousands of speakers, that are described in Section \ref{sssec:librispeech}. Finally, we evaluate the best speaker embedding technique across corpora.

% \subsection{Speaker Attention under Mismatched Condition {on WSJ0-1talker and WSJ0-2talker Datasets} }
\subsection{Speaker Attention to Single- vs. 2-Talker Speech {on WSJ0-1talker and WSJ0-2talker Datasets} }
\label{ssec:eval_extraction}

%\textcolor{red}{which dataset is used for this experiment?} \textcolor{blue}{(I also summarized which corpora are used for each section in the first paragraph in Section IV, as shown in blue color.)} 
While SpEx+ system~\cite{ge2020spex+} demonstrates superior target speaker extraction performance when trained and evaluated on multi-talker speech, it remains unknown how such system performs in face of the single talker speech. To understand the impact of mismatch between training and testing, we use the single-talker speech to test the SpEx+ system that trained solely on the 2-talker mixture data. We refer to the resulted system as SpEx+ I. From Table \ref{tab:spex}, we notice that the SpEx+ I system achieves 18.1 dB and 10.9 dB in terms of SI-SDR for 2-talker and single talker speech, respectively. This huge performance gap can be explained by the mismatch of training and testing conditions. 

%As the single talker speech data is not seen during training, we observe a significant performance drop when the system is tested on single talker speech. 
  
To study how much the multi-condition training, using both single and 2-talker speech data, helps in improving the system generalizability, we fine-tuned the SpEx+ I system using data from both conditions, which we referred to as SpEx+ II. From Table \ref{tab:spex}, we note the performance on single talker speech improved significantly from 10.9 dB to 56.7 dB, while for 2-talker speech is still remains comparable. Motivated by this result, we apply the multi-condition training protocol to construct a unified tSV system that works for both single- and multi-talker speech in this work. 
  
  %Unfortunately, the SpEx+ doesn't work well. We understand that SpEx+ is an end-to-end system that directly takes time-domain signal as inputs. If 2-talker mixture speech and clean single speaker's voice are given to the random initialized model, it may be not easy for the model to be well trained with two different data distributions. 
 
\begin{table}[htb]
    \centering
    \caption{Performance of SpEx+ on WSJ0-2mix-extr database under mismatch conditions between training and evaluation.}
    %\vspace{-6pt}
    \begin{tabular}{|l|c|c|c|} \hline
      \textbf{Systems} & \textbf{Training data} & \textbf{Test data} & \textbf{SI-SDR (dB)} \\ \hline\hline
      \multirow{2}{*}{SpEx+ I} & Mixture & Mixture & 18.1 \\ 
       & Mixture & Single & 10.9 \\ \hline
      \multirow{2}{*}{SpEx+ II} & Mixture+Single & Mixture & 17.9 \\ 
       & Mixture+Single & Single & 56.7 \\ \hline
    \end{tabular}
    \label{tab:spex}
    \vspace{-10pt}
\end{table}

\subsection{Speaker Representation for {SV on WSJ0-1talker Dataset} }
\label{ssec:eval_sv}

\begin{table}[t]
    \centering
    \caption{A comparison between traditional SV speaker embeddings and the speaker embeddings by the proposed speaker representation module in Figure \ref{fig:system_sesv} under single-talker condition on WSJ0-1talker Dataset. }
    %\vspace{-6pt}
    \begin{tabular}{|l|c|c|c|} \hline
      \textbf{Systems} & \textbf{EER (\%)} & \textbf{DCF08} & \textbf{DCF10} \\ \hline\hline
      i-vector PLDA \cite{Dehak&Kenny2011} & 3.00 & \textbf{0.360} & 0.522 \\ 
      x-vector PLDA \cite{snyder2016deep} & 5.87 & 0.692 & 0.888 \\ \hline
      SV-T & 4.40 & 0.450 & 0.630 \\ 
      SV-F & 4.37 & 0.416 & 0.587 \\ 
      SV-FA & \textbf{2.90} & 0.363 & \textbf{0.517} \\ \hline
    \end{tabular}
    \vspace{-10pt}
    \label{tab:SV_traditional}
\end{table}

\begin{table*}[t]
    \centering
    \caption{%Comparisons between tSV systems and traditional SV systems with and without target speaker extraction on WSJ0-2mix-extr corpus.
    A summary of target speaker verification experiments on (a) WSJ0-1taker and WSJ0-2talker, (b) Libri-1talker and Libri-2talker datasets.  ``Path'' denotes the output path of the speaker attention module in Figure \ref{fig:system_sesv}. ``Speech Data Type'' column indicates the type of training and evaluation data involved (`M' for multi-talker, `S' for single-talker).  In ``Speaker Embedding Input'' column,  ``Training'', ``Test'', and ``Enrol" indicate the type of speech representation for speaker embedding, where ``$s$'' and ``$y$" represent single- and 2-talker speech; ``$x$" is the original single-talker enrollment speech; ``$\hat{S}$", ``$\hat{S}_{test}$" and ``$\hat{S}_{ref}$" are the modulated  embedding  coefficients from the speaker attention module; ``$\hat{s}$", ``$\hat{s}_{test}$" and ``$\hat{s}_{ref}$" are the reconstructed signals from the speaker attention module. ``Baseline1'' represents the zero-effort test case where SV system is trained with single-talker speech and evaluated on 2-talker speech. ``Baseline2'' represents the speaker extraction-verification pipeline systems. ``Upper Bound'' denotes the case where single-talker speech is used in both training and evaluation. ``$^{\dagger}$" represents the alternative enrollment process in Figure \ref{fig:system_inference}.}
    %\vspace{-6pt}
    % \scriptsize
    \begin{tabular}{|l|l|c|c|c|c|c|c|c|c|c|c|} \hline
      \multirow{2}{*}{\textbf{System ID}} & \multirow{2}{*}{\textbf{Architecture}} & \multirow{2}{*}{\textbf{Path}} & \multirow{2}{*}{\textbf{Joint Opt.}} & \multicolumn{2}{c|}{\textbf{Speech Data Type}} & \multicolumn{3}{c|}{\textbf{Speaker Embedding Input}} & \multirow{2}{*}{\textbf{EER (\%)}} & \multirow{2}{*}{\textbf{DCF08}} & \multirow{2}{*}{\textbf{DCF10}} \\ \cline{5-9}
       &  &  &  & \textbf{Training} & \textbf{Test} & \textbf{Training} & \textbf{Test} & \textbf{Enrol} &  & & \\ \hline\hline
       \multicolumn{12}{|c|}{(a) Single- and multi-talker speaker verification experiments on WSJ0-1talker and WSJ0-2talker} \\ \hline \hline
      1 (Baseline1) & SV-T & - & - & S & M & $s$ & $y$ & $x$ & 19.37 & 0.864 & 0.928 \\ 
      2 (Baseline2) & TSE-SV-T & - & No & M+S & M & $s$ & $\hat{s}_{test}$ & $x$ & 16.97 & 0.815 & 0.897 \\\hline
      3 & tSV-R & A & Yes & M+S & M & $\hat{S}$ & $\hat{S}_{test}$ & $\hat{S}_{ref}$ & 6.67 & 0.580 & 0.774 \\ 
      4 & tSV-R & A & Yes & M+S & S & $\hat{S}$ & $\hat{S}_{test}$ & $\hat{S}_{ref}$ & 4.00 & 0.371 & 0.561 \\ \hline
      5 & tSV-T & B & Yes & M+S & M & $\hat{s}$ & $\hat{s}_{test}$ & $\hat{s}_{ref}$ & 6.60 & 0.582 & 0.759 \\ 
      6 & tSV-T & B & Yes & M+S & S & $\hat{s}$ & $\hat{s}_{test}$ & $\hat{s}_{ref}$ & 4.63 & 0.424 & 0.671 \\
      7 & tSV-T & B & Yes & M+S & M & $\hat{s}$ & $\hat{s}_{test}$ & $x^{\dagger}$ & 62.13 & 1.000 & 1.000 \\ 
      8 & tSV-T & B & Yes & M+S & S & $\hat{s}$ & $\hat{s}_{test}$ & $x^{\dagger}$ & 77.93 & 1.000 & 1.000 \\ \hline
      9 (Upper Bound) & SV-T & - & - & S & S & $s$ & $s$ & $x$ & 4.40 & 0.450 & 0.630 \\ \hline \hline
      
      10 (Baseline1) & SV-F & - & - & S & M & $s$ & $y$ & $x$ & 20.07 & 0.849 & 0.894 \\
      11 (Baseline2) & TSE-SV-F & - & No & M+S & M & $s$ & $\hat{s}_{test}$ & $x$ & 10.53 & 0.686 & 0.789 \\\hline
      12 & tSV-F & B & Yes & M+S & M & $\hat{s}$ & $\hat{s}_{test}$ & $\hat{s}_{ref}$ & 6.67 & 0.573 & 0.721 \\ 
      13 & tSV-F & B & Yes & M+S & S & $\hat{s}$ & $\hat{s}_{test}$ & $\hat{s}_{ref}$ & 3.87 & 0.352 & 0.531 \\
      14 & tSV-F & B & Yes & M+S & M & $\hat{s}$ & $\hat{s}_{test}$ & $x^{\dagger}$ & 6.53 & 0.570 & 0.708 \\ 
      15 & tSV-F & B & Yes & M+S & S & $\hat{s}$ & $\hat{s}_{test}$ & $x^{\dagger}$ & 4.03 & 0.380 & 0.553 \\ \hline
      16 (Upper Bound) & SV-F & - & - & S & S & $s$ & $s$ & $x$ & 4.37 & 0.416 & 0.587 \\ \hline \hline
      
      17 (Baseline1) & SV-FA & - & - & S & M & $s$ & $y$ & $x$ & 20.97 & 0.825 & 0.893 \\
      18 (Baseline2) & TSE-SV-FA & - & No & M+S & M & $s$ & $\hat{s}_{test}$ & $x$ & 6.63 & 0.583 & 0.759 \\\hline
      19 & tSV-FA & B & Yes & M+S & M & $\hat{s}$ & $\hat{s}_{test}$ & $\hat{s}_{ref}$ & 5.03 & 0.449 & 0.604 \\ 
      20 & tSV-FA & B & Yes & M+S & S & $\hat{s}$ & $\hat{s}_{test}$ & $\hat{s}_{ref}$ & 2.63 & 0.325 & 0.505 \\
      21 & tSV-FA & B & Yes & M+S & M & $\hat{s}$ & $\hat{s}_{test}$ & $x^{\dagger}$ & 5.13 & 0.443 & 0.602 \\
      22 & tSV-FA & B & Yes & M+S & S & $\hat{s}$ & $\hat{s}_{test}$ & $x^{\dagger}$ & 2.73 & 0.325 & 0.492 \\ \hline
      23 (Upper Bound) & SV-FA & - & - & S & S & $s$ & $s$ & $x$ & 2.90 & 0.363 & 0.517 \\ \hline \hline
      \multicolumn{12}{|c|}{(b) Single- and multi-talker speaker verification experiments on Libri-1talker and Libri-2talker} \\ \hline \hline
      24 (Baseline1) & SV-FA & - & - & S & M & $s$ & $y$ & $x$ & 17.62 & 0.855 & 0.938 \\ \hline
      25 & tSV-FA & B & Yes & M+S & M & $\hat{s}$ & $\hat{s}_{test}$ & $\hat{s}_{ref}$ & 7.88 & 0.508 & 0.681 \\ 
      26 & tSV-FA & B & Yes & M+S & S & $\hat{s}$ & $\hat{s}_{test}$ & $\hat{s}_{ref}$  & 6.22 & 0.348 & 0.675 \\
      27 & tSV-FA & B & Yes & M+S & M & $\hat{s}$ & $\hat{s}_{test}$ & $x^{\dagger}$ &  7.80 & 0.514 & 0.685 \\
      28 & tSV-FA & B & Yes & M+S & S & $\hat{s}$ & $\hat{s}_{test}$ & $x^{\dagger}$ & 6.33 & 0.354 & 0.677 \\ \hline
      29 (Upper Bound) & SV-FA & - & - & S & S & $s$ & $s$ & $x$ & 5.70 & 0.293 & 0.403 \\ \hline
    \end{tabular}
    \vspace{-10pt}
    \label{tab:SV_overall}
\end{table*}

%\textcolor{red}{which dataset is used for this experiment?} \textcolor{blue}{(I also summarized which corpora are used for each section in the first paragraph in Section IV, as shown in blue color.)}

{The speaker representation module in Figure \ref{fig:system_sesv} is designed to work with the speaker attention module for target speaker recognition. Therefore, the speaker embedding process in tSV is different from that in other traditional speaker verification systems where single talker speech is assumed. To justify the effectiveness of proposed speaker representation module, we compare various speaker embedding SV systems, namely, time-domain SV (SV-T), the frequency SV (SV-F) and the frequency-domain attention-based SV (SV-FA), as shown in Figure \ref{fig:speaker_embedding}(b), \ref{fig:speaker_embedding}(c) and \ref{fig:speaker_embedding}(d) respectively, with the traditional single-talker systems, i.e., x-vector PLDA \cite{snyder2016deep}, under the single talker test condition. The traditional SV system with the single-talker assumption is also referred to as the single-talker SV system. In this study, the speaker attention module is not required for our system.}

{We observe from Table \ref{tab:SV_traditional} that 1)  SV-T, SV-F, and SV-FA systems consistently outperform the x-vector PLDA system in terms of EER; 2)  SV-FA system achieves the best EER among all systems. The results suggest that the proposed speaker representation module is as competitive as, if not better than, i-vector or x-vector systems for single talker speech. We proceed with our proposed speaker representation module in the subsequent tSV studies.}

\subsection{Evaluating Target Speaker Embeddings on WSJ0-2talker Dataset}
\label{ssec:eval_tSV}

%\textcolor{blue}{(Chenglin explains: In this section, we mainly reported the performances of the four tSV systems on 2-talker speech. Since we need to compare with the baseline and upperbound, the performances of the traditional SV trained with sing-talker speech are included in this section by evaluating on 2-talker speech (Baseline1) and single-talker speech (upperbound). The performance of tSV systems on single-talker speech is reported in next section, not here.)}

%\textcolor{red}{which dataset is used for this experiment?} \textcolor{blue}{(I also summarized which corpora are used for each section in the first paragraph in Section IV, as shown in blue color.)} 
To investigate the effect of multi-talker speech on single-talker SV, we compare the traditional SV and the proposed tSV with four different target speaker embedding schemes, namely tSV-R, tSV-T, tSV-F, and tSV-FA, on 2-talker mixture evaluation dataset, i.e. WSJ0-2talker.  

The system setups and experiment results are summarised in Table~\ref{tab:SV_overall} (a). First, we report three zero-effort baselines (System 1, 10, and 17) that follow traditional speaker embedding schemes, SV-T, SV-F, and  SV-FA. They are trained on single talker dataset (WSJ0-1talker) and evaluated on 2-talker mixture evaluation set. We further report 3 reference systems (System 9, 16, and 23) that are trained and evaluated both on single-talker speech. As they don't include interference speakers, the speaker attention module is not involved in the study. For multi-talker experiments, System 9 represents the upper-bound performance of tSV-R and tSV-T; Systems 16 and 23 represent the upper-bound performances of tSV-F and tSV-FA respectively.

For the upper-bound reference systems, we observe that the SV-FA system outperforms the SV-T and SV-F systems, benefiting from the dynamic features and attention pooling. Comparing System 1 and 9, 10 and 16, and 17 and 23, we observe that the performance of the single-talker SV systems seriously degrade in the presence of interference speakers.

%System 9 of Table \ref{tab:SV_overall} (a) shows the upper-bound performance of tSV-R and tSV-T methods for overlapped multi-talker SV. This upper-bound performance is evaluated on clean single speaker's voice (upper-bound of the extracted speech) with the model trained on the same condition.

We also report three competitive baselines (System 2, 11, and 18) that follow the target speaker extraction-verification (TSE-SV) pipeline~\cite{rao2019target_is}, where speaker extraction and speaker verification modules are trained separately. Between the zero-effort baselines and competitive baselines, in particular, between System 17 and 18, we observe the followings: (1) The target speaker extraction front-end greatly improves the SV performance under multi-talker test condition; %(2) Comparing between System 17 and System 18, the TSE-SV-FA system achieves 68.4\%, 29.3\% and 15.0\% relative improvements in terms of EER, DCF08 and DCF10. 
(2) Among System 2, 11 and 18, the frequency-domain SV systems (SV-F and SV-FA) appear to be more robust than the time-domain counterpart (SV-T). {This suggests that the time-domain SV system is sensitive to the mismatch between speaker verification module trained with single talker speech, and the extracted speech from speaker extraction module.}

The difference between the target speaker extraction-verification pipeline \cite{rao2019target_is} and the proposed tSV system mainly lies in their training schemes. The former trains the extraction module and verification module separately, while the latter enables a multi-task joint training between speaker attention and speaker representation modules. Next we summarize the results for the four target speaker embedding schemes under multi-talker test condition. 

%The tSV-R system follows \textit{Path A}, while tSV-T, tSV-F and tSV-FA systems follow \textit{Path B} as shown in Figure \ref{fig:system_sesv}. 
Now we look into 4 groups of experiments for 4 target speaker embedding schemes in Table \ref{tab:SV_overall} (a). Let's examine how the systems perform on 2-talker speech including System 3 (tSV-R), Systems 5 (tSV-T), Systems 12 (tSV-F), and Systems 19 (tSV-FA). We observe the followings: (1) tSV-FA achieves significantly better performance than other three schemes, which deliver comparable results. (2) tSV-FA improves the performance on 2-talker speech by 76.0\%, 45.6\% and 32.4\% relative improvement over the single talker trained SV baseline (System 17) in terms of EER, DCF08 and DCF10, respectively. (3) %Comparing with TSE-SV-FA (System 18) without joint optimization, 
tSV-FA achieves 24.1\%, 23.0\% and 20.4\% relative improvement over the competitive baseline (System 18) in terms of EER, DCF08 and DCF10, respectively, that is attributed to the joint optimization between the speaker attention and speaker representation modules. %(4) Although tSV-R, tSV-T and tSV-F methods are no better than tSV-FA, these three methods also achieve dramatically improvement comparing with the corresponding baseline systems. 
(4) tSV with all four target speaker embedding schemes consistently outperform zero-effort baselines and competitive baselines for multi-talker speech.  

It is encouraging to see that the proposed tSV systems evaluated on multi-talker speech achieve comparable performance with the upper-bound reference systems evaluated on single speaker speech.

\subsection{Evaluating Target Speaker Embeddings on WSJ0-1talker Dataset}
\label{ssec:eval_tSVf}

In real world applications, one expects that the same speaker verification system is able to handle single talker and multi-talker speech seamlessly because we don't know in advance whether the speakers would overlap. We have studied the performance of tSV systems for multi-talker speech at different overlapping rate in Section \ref{ssec:eval_tSV}.  It is interesting to know the performance of the proposed tSV systems on single talker's evaluation set, i.e. with zero overlapping rate. The results are summarized as System 4, 6, 13 and 20 in Table \ref{tab:SV_overall} (a).

To evaluate under single talker condition, we adopt the traditional single talker SV system as a baseline. Comparing among Systems 4, 6 and 9, we observe that the tSV-R system (System 4) achieves significant better performance than SV-T system (System 9) in terms of EER, DCF08 and DCF10. While this is a pleasantly surprising result, we consider that the target speaker extraction system in tSV-R does the first-round verification by only extracting the target speaker's voice. This extra step helps to decline the non-target trials. Although the tSV-T system only shows superior performance in terms of DCF08, we observe the same findings as tSV-R system in both tSV-F and tSV-FA systems (System 13 vs. 16 and System 20 vs. 23) in terms of EER, DCF08 and DCF10. With the same single talker evaluation condition, the tSV-FA method achieves the best performance of 2.63\% (EER), 0.325 (DCF08) and 0.505 (DCF10) among the proposed four approaches. Comparing with the SV-FA system (System 23) trained and evaluated on same clean single speaker condition, the tSV-FA method achieves 9.3\%, 10.5\% and 2.3\% relative improvements in terms of EER, DCF08 and DCF10.

From Section \ref{ssec:eval_tSV}, we conclude that the proposed tSV systems significantly improve multi-talker speaker verification. When the same model is evaluated in this section on single talker speech, they even outperform the traditional SV systems trained and evaluated for single talker condition. With these findings, we confirm that the proposed tSV system with various target speaker embedding schemes represents an unified framework for single and multi-talker speech, with tSV-FA as the top achiever.

\subsection{Alternative Configuration for Target Speaker Enrollment}
\label{ssec:eval_enrol}

As enrollment utterance always contains single talker's voice, Figure \ref{fig:system_inference} shows an alternative to the standard configuration in Figure \ref{fig:system_sesv}. In the alternative configuration, we directly extract $e_{ref}$ from the enrollment utterance $x(t)$ bypassing the speaker attention module. This alternative configuration works for all \textit{Path B} schemes, i.e., tSV-T, tSV-F and tSV-FA.

In Table \ref{tab:SV_overall} (a), the systems that follow the alternative configuration in Figure \ref{fig:system_inference} are marked with ``$^\dagger$".  We observe that alternative tSV-F and tSV-FA systems (system 14, 15, 21, and 22) achieve comparable performance to those with the standard configurations (System 12, 13, 19 and 20). This suggests that the extracted speech from the speaker attention module is of similar quality as the original signal as far as speaker embedding is concerned. The SI-SDR performance of the extracted speech (56.7dB) leads to the same conclusion, as reported in Table \ref{tab:spex}.

However, we also observe that the alternative tSV-T system fails short of the expectation when comparing with the standard configuration, as reported in System 7 and 8 of Table \ref{tab:SV_overall} (a). We consider that the speaker representation module adopts a time-domain trainable speech encoder that is accustomed to extracted speech from speaker attention module. Therefore, there is a mismatch between the extracted speech and the original clean speech, that adversely affect the SV performance. This corroborates that the time-domain speaker representation module is sensitive to the mismatch between the extracted speech for training and clean speech for inference.

\subsection{Benchmarking against the State-of-the-art on {WSJ0-2talker Dataset}}
\label{ssec:eval_comp_other}

\begin{table}[t]
    \centering
    \caption{Comparisons between tSV systems and other state-of-the-art systems on WSJ0-2talker dataset. ``OSD-SV'' represents the case where we replace the speaker attention module in Figure~\ref{fig:system_sesv} with an oracle speaker diarization (OSD). ``SE-SV1'' and ``SE-SV2'' are two speaker extraction-verification pipeline systems with SBF-MTSAL and SBF-MTSAL-Concat~\cite{xu2019optimization} as the speaker extraction front-end, respectively.}
    %\vspace{-6pt}
    %\footnotesize
    \begin{tabular}{|l|c|c|c|} \hline
      \textbf{Systems} & \textbf{EER (\%)} & \textbf{DCF08} & \textbf{DCF10} \\ \hline\hline
      OSD-SV & 14.60 & 0.851 & 0.908 \\ 
      SE-SV1 \cite{rao2019target_is} & 8.30 & 0.643 & 0.777 \\
      SE-SV2 \cite{rao2019target_is} & 7.77 & 0.631 & 0.747 \\ \hline
      tSV-R & 6.67 & 0.580 & 0.774 \\
      tSV-T & 6.60 & 0.582 & 0.759 \\ 
      tSV-F & 6.53 & 0.570 & 0.708 \\ 
      tSV-FA & \textbf{5.03} & \textbf{0.449} & \textbf{0.604} \\ \hline 
      Upper-Bound & 2.90 & 0.363 & 0.517 \\ \hline
    \end{tabular}
    \vspace{-10pt}
    \label{tab:SV_comp}
\end{table}

Recently studies, such as personal VAD \cite{ding2019personal}, target VAD \cite{medennikov2020target}, seek to address speaker diarization problem for multi-talker speech. They could be used as the front-end for multi-talker SV when the speakers don't overlap. We are interested in the performance of multi-talker SV with such an oracle speaker diarization front-end. %, which is the upper-bound of those speaker diarization systems with persoanl VAD or targt VAD.
We obtain the ground-truth VAD by firstly applying the energy-based VAD on the single talker speech that makes up the mixture speech, and use the VAD labels as the diarization label for mixture speech.

Since the speech in WSJ0 corpus is quite clean and the energy-based VAD works well on the clean speech, we consider the obtained diarization labels as oracle diarization labels. With the oracle speaker diarization, the average percentage of removed non-target speech frames in the mixture speeches is around $24\%$. ``OSD-SV'' in Table~\ref{tab:SV_comp} shows the performance of multi-talker SV system with the oracle speaker diarization front-end, which also means the best performance achieved by speaker diarization for multi-talker SV. Compared with the multi-talker speaker verification results by tSV-R, tSV-T, tSV-F and tSV-FA, as reported from Table \ref{tab:SV_overall} (a), it is shown that the proposed tSV system significantly outperforms ``OSD-SV''.

We further compare the proposed tSV system with the speaker extraction-verification pipeline, as summarized in Table \ref{tab:SV_comp}. ``SE-SV1'' and ``SE-SV2'' are speaker extraction-verification pipeline systems \cite{rao2019target_is}, where SBF-MTSAL and SBF-MTSAL-Concat methods \cite{xu2019optimization} serve as the speaker extraction front-end, and i-vector PLDA serves as the speaker verification back-end, without joint optimization. We observe the followings: (1) The proposed tSV system with various target speaker embedding schemes consistently outperform SE-SV1 \cite{rao2019target_is} and SE-SV2 \cite{rao2019target_is}; (2) The tSV-FA system achieves 35.3\%, 28.8\% and 19.1\% relative improvement over SE-SV2, the most competitive system, in terms of EER, DCF08 and DCF10, respectively.

\subsection{Evaluating tSV-FA on {Libri-1talker/Libri-2talker Datasets}}
\label{ssec:eval_best_libri2mix}

We further conduct experiments on a large Libri2Mix corpus with 1,172 speakers in the training set. As the tSV-FA system achieves the best performance on WSJ0-1talker and WSJ0-2talker datasets, we only train and evaluate the tSV-FA system on the Libri-1talker and Libri-2talker datasets.  

As summarized in Table \ref{tab:SV_overall} (b), System 24 and 29 are the baseline and the upper-bound systems for multi-talker speaker verification. Both are traditional SV systems trained under single talker condition, while the baseline is evaluated on 2-talker speech, and the upper bound system is evaluated on single talker speech.

 We also evaluate the trained tSV-FA model on both 2-talker and single talker speech, denoted as Systems 25 and 26, just like in Section \ref{ssec:eval_tSV} and Section \ref{ssec:eval_tSVf}. We observe that the tSV-FA system (System 25) achieves 55.3\%, 40.6\% and 27.4\% relative improvements over the baseline (System 24) in terms of EER, DCF08 and DCF10, respectively.
 The performance of tSV-FA system, evaluated either on 2-talker speech (System 25) or on single talker speech (System 26) is approaching that of the upper-bound system (System 29). The results further validate that tSV-FA system represents a unified solution for both single and multi-talker speech.

We further evaluate the alternative configurations in Figure  \ref{fig:system_inference}. Bypassing the speaker attention module, we  directly extract $e_{ref}$ from the enrollment utterance $x(t)$, and evaluate such an alternative tSV-FA system on the same 2-talker (System 27) and single talker (System 28) datasets.  We observe that the alternative systems perform similarly to the standard configurations (Systems 25 and 26).

\subsection{Evaluating tSV-FA across Corpora}
\label{ssec:eval_across}

\begin{table}[t]
    \centering
    \caption{Evaluation of the tSV-FA system across corpora.}
    %\vspace{-6pt}
    % \scriptsize
    \begin{tabular}{|c|c|c|c|c|} \hline
      \textbf{Training Set} & \textbf{Test Set} & \textbf{EER (\%)} & \textbf{DCF08} & \textbf{DCF10} \\ \hline\hline
       & WSJ0-2talker & 5.03 & 0.449 & 0.604 \\ 
      WSJ0-1talker \& & WSJ0-1talker & 2.63 & 0.325 & 0.505 \\ \cline{2-5}
      WSJ0-2talker& Libri-2talker & 16.85 & 0.936 & 0.988 \\
      & Libri-1talker & 11.33 & 0.744 & 0.907 \\ \hline \hline
       & WSJ0-2talker &  4.9 & 0.444 & 0.558 \\ 
      Libri-1talker \& & WSJ0-1talker & 2.43 & 0.322 & 0.467 \\ \cline{2-5}
      Libri-2talker& Libri-2talker & 7.88 & 0.508 & 0.681 \\
      & Libri-1talker & 6.22 & 0.348 & 0.675 \\ \hline
    \end{tabular}
    \vspace{-10pt}
    \label{tab:SV_cross}
\end{table}

We further evaluate tSV-FA, that is trained on either WSJ0 or LibriSpeech dataset, but evaluated across WSJ0 and LibriSpeech datasets. The results are summarized in Table \ref{tab:SV_cross}.

Among the training-evaluation data pairs, we observe a low EER for WSJ0-WSJ0, and LibriSpeech-WSJ0 pairs. The EER for LibriSpeech-WSJ0 is even lower than that for WSJ0-WSJ0, which we consider is due to the fact that LibriSpeech is a larger database with many more speakers, thus, leading to a more robust tSV-FA system. This is an encouraging result,  that confirms the ability of cross-corpus generalization from LibriSpeech-trained model to WSJ0 evaluation condition.  

However, we haven't observed the same generalization ability for WSJ0-trained tSV-FA model. The EER for WSJ0-LibriSpeech is higher than that of LibriSpeech-LibriSpeech.  Nonetheless, these results match our expectation, as we know that system performance under mismatched training-evaluation conditions is always poorer than that under matched conditions. We understand that WSJ0 is a smaller database with a small number of speakers. Furthermore, WSJ0 is recorded in a quiet acoustic environment with only one type of microphone channel, that may leads to the poor generalization under the LibriSpeech test condition. 

%When we increase the number of speakers and microphone channels as well as the recording environments, i.e. Libri-1talker and Libri-2talker, we observe that the tSV-FA method trained on Libri-1talker and Libri-2talker datasets could achieve comparable performance on WSJ0-1talker and WSJ0-2talker evaluation set as the same model trained on the WSJ0-1talker and WSJ0-2talker training sets. With large amount of training data with thousands of speakers and various of microphone channels, the tSV-FA method could be robust and generalized well to unknown speakers and microphone channels.
 
%\textcolor{blue}{(Chenglin: you mean the performances of Libri-2talker (7.88) and Libri-1talker (6.22) evaluation sets are worse than the WSJ0-2talker (4.9) and WSJ0-1talker (2.43) evaluation sets, right? Because the evaluation trials for WSJ and Libri are different. They can't be compared directly. The durations of Libri-2talker and Libri-1talker are short. The recording environment of LibriSpeech is not as clean as WSJ. These are the reasons why the Libri matching conditions are worse than WSJ evaluation sets.)}

\section{Conclusions and Discussions}
\label{sec:con}
%\vspace{-2pt}

In this paper, we present a unified target speaker verification framework tSV for both single- and multi-talker speech. This framework jointly optimizes a speaker attention module and a speaker representation module via a multi-task learning framework. We systematically evaluated the performance of each individual module as well as the whole framework using two standard corpora. Experimental results show that the proposed framework significantly improves speaker verification performance for multi-talker speech, approaching that of traditional SV systems under the single-talker condition. It also outperforms other solutions, for instance, combining the state-of-the-art speaker extraction and speaker verification systems in tandem. Through multi-condition training, using both single- and multi-talker speech, we show that the proposed framework can easily generalize across different testing conditions. 

There have been many studies on speaker extraction and speaker verification in literature. While these studies are usually carried out in isolation. For the first time, we show that with an advanced speaker attention-verification mechanism, we can achieve accurate speaker verification in multi-talker environment using the same enrollment utterance as what we have in the conventional single-talker speaker verification system. 

%The significance of this work lies in the idea of applying selective auditory attention in both speaker extraction and speaker verification. 
%By dropping the single-talker assumption adopted in most of the traditional speaker verification studies, this framework greatly improves the usability of speaker verification system in real world environments. 

%However, we have not evaluated the framework in adverse real-world acoustic environment with noise and reverberations. 
We would like to acknowledge that preparing a system to work in an unknown real-world acoustic environment remains a challenge. It is not the purpose of this work to propose a universal solution. Instead, we have devoted this work to the exploration of a unified speaker verification solution for both single and multi-talker speech. Although we only conducted experiments on single and 2-talker speech, the proposed framework can be easily extended to more complex mixtures that have three or more speakers, together with noise and reverberation. As future work, we are particularly interested in a training strategy, that can be enabled by self-supervised learning~\cite{selfsupervised} and incremental learning \cite{wu2019large}, to allow the tSV system to adapt itself in the real-world changing environments.% We will also evaluate the proposed framework on public domain speaker verification corpus.

% or


%\appendices

%\section{}

% % use section* for acknowledgment
% \section*{Acknowledgment}
% To be added.

\ifCLASSOPTIONcaptionsoff
  \newpage
\fi



\bibliographystyle{IEEEtran}
\bibliography{2020taslp}

\end{document}